\documentclass[12pt,prd,tightenlines,nofootinbib,showpacs,showkeys]{revtex4}
\newcommand{\be}{\begin{equation}}
\newcommand{\ee}{\end{equation}}
\usepackage{bm}
\usepackage{graphics}
\usepackage{rotating}
\usepackage{epsfig}
\begin{document}
\title{\begin{flushright}{\rm\normalsize HU-EP-06/13}\end{flushright}
Relativistic effects in the production of pseudoscalar and vector
doubly heavy mesons from $e^+e^-$ annihilation}
\author{D.Ebert\footnote{E-mail:~debert@physik.hu-berlin.de}}
\affiliation{Institut f\"ur Physik, Humboldt--Universit\"at zu Berlin,
Newtonstr. 15, D-12489  Berlin, Germany}
\author{A. P.
Martynenko\footnote{E-mail:~apm@physik.hu-berlin.de;~mart@ssu.samara.ru}}
\affiliation{Institut f\"ur Physik, Humboldt--Universit\"at zu Berlin,
Newtonstr. 15, D-12489  Berlin, Germany}
\affiliation{Samara State University, Pavlov Street 1, Samara 443011,
Russia}

\begin{abstract}
On the basis of the perturbative QCD and the
relativistic quark model we investigate the
relativistic and bound state effects in the production processes of a pair of
$S$-wave doubly heavy mesons with opposite charge conjugation
consisting of $b$ and $c$ quarks. All possible relativistic
corrections in the production amplitude including the terms
connected with the transformation law of the bound state wave
function to the reference frame of the moving pseudoscalar ${\cal P}-$ and
vector ${\cal V}-$ mesons are taken into account. We obtain a growth of
the cross section for the reaction $e^++e^-\to J/\Psi+ \eta_c$ due to considered effects by
a factor $2\div 2.5$ in the range of the center-of-mass energy
$\sqrt{s}=6\div 12$ GeV.
\end{abstract}



\maketitle

\immediate\write16{<<WARNING: LINEDRAW macros work with emTeX-dvivers
                    and other drivers supporting emTeX \special's
                    (dviscr, dvihplj, dvidot, dvips, dviwin, etc.) >>}

\newdimen\Lengthunit       \Lengthunit  = 1.5cm
\newcount\Nhalfperiods     \Nhalfperiods= 9
\newcount\magnitude        \magnitude = 1000

\catcode`\*=11
\newdimen\L*   \newdimen\d*   \newdimen\d**
\newdimen\dm*  \newdimen\dd*  \newdimen\dt*
\newdimen\a*   \newdimen\b*   \newdimen\c*
\newdimen\a**  \newdimen\b**
\newdimen\xL*  \newdimen\yL*
\newdimen\rx*  \newdimen\ry*
\newdimen\tmp* \newdimen\linwid*

\newcount\k*   \newcount\l*   \newcount\m*
\newcount\k**  \newcount\l**  \newcount\m**
\newcount\n*   \newcount\dn*  \newcount\r*
\newcount\N*   \newcount\*one \newcount\*two  \*one=1 \*two=2
\newcount\*ths \*ths=1000
\newcount\angle*  \newcount\q*  \newcount\q**
\newcount\angle** \angle**=0
\newcount\sc*     \sc*=0

\newtoks\cos*  \cos*={1}
\newtoks\sin*  \sin*={0}

\catcode`\[=13

\def\rotate(#1){\advance\angle**#1\angle*=\angle**
\q**=\angle*\ifnum\q**<0\q**=-\q**\fi
\ifnum\q**>360\q*=\angle*\divide\q*360\multiply\q*360\advance\angle*-\q*\fi
\ifnum\angle*<0\advance\angle*360\fi\q**=\angle*\divide\q**90\q**=\q**
\def\sgcos*{+}\def\sgsin*{+}\relax
\ifcase\q**\or
 \def\sgcos*{-}\def\sgsin*{+}\or
 \def\sgcos*{-}\def\sgsin*{-}\or
 \def\sgcos*{+}\def\sgsin*{-}\else\fi
\q*=\q**
\multiply\q*90\advance\angle*-\q*
\ifnum\angle*>45\sc*=1\angle*=-\angle*\advance\angle*90\else\sc*=0\fi
\def[##1,##2]{\ifnum\sc*=0\relax
\edef\cs*{\sgcos*.##1}\edef\sn*{\sgsin*.##2}\ifcase\q**\or
 \edef\cs*{\sgcos*.##2}\edef\sn*{\sgsin*.##1}\or
 \edef\cs*{\sgcos*.##1}\edef\sn*{\sgsin*.##2}\or
 \edef\cs*{\sgcos*.##2}\edef\sn*{\sgsin*.##1}\else\fi\else
\edef\cs*{\sgcos*.##2}\edef\sn*{\sgsin*.##1}\ifcase\q**\or
 \edef\cs*{\sgcos*.##1}\edef\sn*{\sgsin*.##2}\or
 \edef\cs*{\sgcos*.##2}\edef\sn*{\sgsin*.##1}\or
 \edef\cs*{\sgcos*.##1}\edef\sn*{\sgsin*.##2}\else\fi\fi
\cos*={\cs*}\sin*={\sn*}\global\edef\gcos*{\cs*}\global\edef\gsin*{\sn*}}\relax
\ifcase\angle*[9999,0]\or
[999,017]\or[999,034]\or[998,052]\or[997,069]\or[996,087]\or
[994,104]\or[992,121]\or[990,139]\or[987,156]\or[984,173]\or
[981,190]\or[978,207]\or[974,224]\or[970,241]\or[965,258]\or
[961,275]\or[956,292]\or[951,309]\or[945,325]\or[939,342]\or
[933,358]\or[927,374]\or[920,390]\or[913,406]\or[906,422]\or
[898,438]\or[891,453]\or[882,469]\or[874,484]\or[866,499]\or
[857,515]\or[848,529]\or[838,544]\or[829,559]\or[819,573]\or
[809,587]\or[798,601]\or[788,615]\or[777,629]\or[766,642]\or
[754,656]\or[743,669]\or[731,681]\or[719,694]\or[707,707]\or
\else[9999,0]\fi}

\catcode`\[=12

\def\GRAPH(hsize=#1)#2{\hbox to #1\Lengthunit{#2\hss}}

\def\Linewidth#1{\global\linwid*=#1\relax
\global\divide\linwid*10\global\multiply\linwid*\mag
\global\divide\linwid*100\special{em:linewidth \the\linwid*}}

\Linewidth{.4pt}
\def\sm*{\special{em:moveto}}
\def\sl*{\special{em:lineto}}
\let\moveto=\sm*
\let\lineto=\sl*
\newbox\spm*   \newbox\spl*
\setbox\spm*\hbox{\sm*}
\setbox\spl*\hbox{\sl*}

\def\mov#1(#2,#3)#4{\rlap{\L*=#1\Lengthunit
\xL*=#2\L* \yL*=#3\L*
\xL*=\xscale\xL* \yL*=\yscale\yL*
\rx* \the\cos*\xL* \tmp* \the\sin*\yL* \advance\rx*-\tmp*
\ry* \the\cos*\yL* \tmp* \the\sin*\xL* \advance\ry*\tmp*
\kern\rx*\raise\ry*\hbox{#4}}}

\def\rmov*(#1,#2)#3{\rlap{\xL*=#1\yL*=#2\relax
\rx* \the\cos*\xL* \tmp* \the\sin*\yL* \advance\rx*-\tmp*
\ry* \the\cos*\yL* \tmp* \the\sin*\xL* \advance\ry*\tmp*
\kern\rx*\raise\ry*\hbox{#3}}}

\def\lin#1(#2,#3){\rlap{\sm*\mov#1(#2,#3){\sl*}}}

\def\arr*(#1,#2,#3){\rmov*(#1\dd*,#1\dt*){\sm*
\rmov*(#2\dd*,#2\dt*){\rmov*(#3\dt*,-#3\dd*){\sl*}}\sm*
\rmov*(#2\dd*,#2\dt*){\rmov*(-#3\dt*,#3\dd*){\sl*}}}}

\def\arrow#1(#2,#3){\rlap{\lin#1(#2,#3)\mov#1(#2,#3){\relax
\d**=-.012\Lengthunit\dd*=#2\d**\dt*=#3\d**
\arr*(1,10,4)\arr*(3,8,4)\arr*(4.8,4.2,3)}}}

\def\arrlin#1(#2,#3){\rlap{\L*=#1\Lengthunit\L*=.5\L*
\lin#1(#2,#3)\rmov*(#2\L*,#3\L*){\arrow.1(#2,#3)}}}

\def\dasharrow#1(#2,#3){\rlap{{\Lengthunit=0.9\Lengthunit
\dashlin#1(#2,#3)\mov#1(#2,#3){\sm*}}\mov#1(#2,#3){\sl*
\d**=-.012\Lengthunit\dd*=#2\d**\dt*=#3\d**
\arr*(1,10,4)\arr*(3,8,4)\arr*(4.8,4.2,3)}}}

\def\clap#1{\hbox to 0pt{\hss #1\hss}}

\def\ind(#1,#2)#3{\rlap{\L*=.1\Lengthunit
\xL*=#1\L* \yL*=#2\L*
\rx* \the\cos*\xL* \tmp* \the\sin*\yL* \advance\rx*-\tmp*
\ry* \the\cos*\yL* \tmp* \the\sin*\xL* \advance\ry*\tmp*
\kern\rx*\raise\ry*\hbox{\lower2pt\clap{$#3$}}}}

\def\sh*(#1,#2)#3{\rlap{\dm*=\the\n*\d**
\xL*=\xscale\dm* \yL*=\yscale\dm* \xL*=#1\xL* \yL*=#2\yL*
\rx* \the\cos*\xL* \tmp* \the\sin*\yL* \advance\rx*-\tmp*
\ry* \the\cos*\yL* \tmp* \the\sin*\xL* \advance\ry*\tmp*
\kern\rx*\raise\ry*\hbox{#3}}}

\def\calcnum*#1(#2,#3){\a*=1000sp\b*=1000sp\a*=#2\a*\b*=#3\b*
\ifdim\a*<0pt\a*-\a*\fi\ifdim\b*<0pt\b*-\b*\fi
\ifdim\a*>\b*\c*=.96\a*\advance\c*.4\b*
\else\c*=.96\b*\advance\c*.4\a*\fi
\k*\a*\multiply\k*\k*\l*\b*\multiply\l*\l*
\m*\k*\advance\m*\l*\n*\c*\r*\n*\multiply\n*\n*
\dn*\m*\advance\dn*-\n*\divide\dn*2\divide\dn*\r*
\advance\r*\dn*
\c*=\the\Nhalfperiods5sp\c*=#1\c*\ifdim\c*<0pt\c*-\c*\fi
\multiply\c*\r*\N*\c*\divide\N*10000}

\def\dashlin#1(#2,#3){\rlap{\calcnum*#1(#2,#3)\relax
\d**=#1\Lengthunit\ifdim\d**<0pt\d**-\d**\fi
\divide\N*2\multiply\N*2\advance\N*\*one
\divide\d**\N*\sm*\n*\*one\sh*(#2,#3){\sl*}\loop
\advance\n*\*one\sh*(#2,#3){\sm*}\advance\n*\*one
\sh*(#2,#3){\sl*}\ifnum\n*<\N*\repeat}}

\def\dashdotlin#1(#2,#3){\rlap{\calcnum*#1(#2,#3)\relax
\d**=#1\Lengthunit\ifdim\d**<0pt\d**-\d**\fi
\divide\N*2\multiply\N*2\advance\N*1\multiply\N*2\relax
\divide\d**\N*\sm*\n*\*two\sh*(#2,#3){\sl*}\loop
\advance\n*\*one\sh*(#2,#3){\kern-1.48pt\lower.5pt\hbox{\rm.}}\relax
\advance\n*\*one\sh*(#2,#3){\sm*}\advance\n*\*two
\sh*(#2,#3){\sl*}\ifnum\n*<\N*\repeat}}

\def\shl*(#1,#2)#3{\kern#1#3\lower#2#3\hbox{\unhcopy\spl*}}

\def\trianglin#1(#2,#3){\rlap{\toks0={#2}\toks1={#3}\calcnum*#1(#2,#3)\relax
\dd*=.57\Lengthunit\dd*=#1\dd*\divide\dd*\N*
\divide\dd*\*ths \multiply\dd*\magnitude
\d**=#1\Lengthunit\ifdim\d**<0pt\d**-\d**\fi
\multiply\N*2\divide\d**\N*\sm*\n*\*one\loop
\shl**{\dd*}\dd*-\dd*\advance\n*2\relax
\ifnum\n*<\N*\repeat\n*\N*\shl**{0pt}}}

\def\wavelin#1(#2,#3){\rlap{\toks0={#2}\toks1={#3}\calcnum*#1(#2,#3)\relax
\dd*=.23\Lengthunit\dd*=#1\dd*\divide\dd*\N*
\divide\dd*\*ths \multiply\dd*\magnitude
\d**=#1\Lengthunit\ifdim\d**<0pt\d**-\d**\fi
\multiply\N*4\divide\d**\N*\sm*\n*\*one\loop
\shl**{\dd*}\dt*=1.3\dd*\advance\n*\*one
\shl**{\dt*}\advance\n*\*one
\shl**{\dd*}\advance\n*\*two
\dd*-\dd*\ifnum\n*<\N*\repeat\n*\N*\shl**{0pt}}}

\def\w*lin(#1,#2){\rlap{\toks0={#1}\toks1={#2}\d**=\Lengthunit\dd*=-.12\d**
\divide\dd*\*ths \multiply\dd*\magnitude
\N*8\divide\d**\N*\sm*\n*\*one\loop
\shl**{\dd*}\dt*=1.3\dd*\advance\n*\*one
\shl**{\dt*}\advance\n*\*one
\shl**{\dd*}\advance\n*\*one
\shl**{0pt}\dd*-\dd*\advance\n*1\ifnum\n*<\N*\repeat}}

\def\l*arc(#1,#2)[#3][#4]{\rlap{\toks0={#1}\toks1={#2}\d**=\Lengthunit
\dd*=#3.037\d**\dd*=#4\dd*\dt*=#3.049\d**\dt*=#4\dt*\ifdim\d**>10mm\relax
\d**=.25\d**\n*\*one\shl**{-\dd*}\n*\*two\shl**{-\dt*}\n*3\relax
\shl**{-\dd*}\n*4\relax\shl**{0pt}\else
\ifdim\d**>5mm\d**=.5\d**\n*\*one\shl**{-\dt*}\n*\*two
\shl**{0pt}\else\n*\*one\shl**{0pt}\fi\fi}}

\def\d*arc(#1,#2)[#3][#4]{\rlap{\toks0={#1}\toks1={#2}\d**=\Lengthunit
\dd*=#3.037\d**\dd*=#4\dd*\d**=.25\d**\sm*\n*\*one\shl**{-\dd*}\relax
\n*3\relax\sh*(#1,#2){\xL*=\xscale\dd*\yL*=\yscale\dd*
\kern#2\xL*\lower#1\yL*\hbox{\sm*}}\n*4\relax\shl**{0pt}}}

\def\shl**#1{\c*=\the\n*\d**\d*=#1\relax
\a*=\the\toks0\c*\b*=\the\toks1\d*\advance\a*-\b*
\b*=\the\toks1\c*\d*=\the\toks0\d*\advance\b*\d*
\a*=\xscale\a*\b*=\yscale\b*
\rx* \the\cos*\a* \tmp* \the\sin*\b* \advance\rx*-\tmp*
\ry* \the\cos*\b* \tmp* \the\sin*\a* \advance\ry*\tmp*
\raise\ry*\rlap{\kern\rx*\unhcopy\spl*}}

\def\wlin*#1(#2,#3)[#4]{\rlap{\toks0={#2}\toks1={#3}\relax
\c*=#1\l*\c*\c*=.01\Lengthunit\m*\c*\divide\l*\m*
\c*=\the\Nhalfperiods5sp\multiply\c*\l*\N*\c*\divide\N*\*ths
\divide\N*2\multiply\N*2\advance\N*\*one
\dd*=.002\Lengthunit\dd*=#4\dd*\multiply\dd*\l*\divide\dd*\N*
\divide\dd*\*ths \multiply\dd*\magnitude
\d**=#1\multiply\N*4\divide\d**\N*\sm*\n*\*one\loop
\shl**{\dd*}\dt*=1.3\dd*\advance\n*\*one
\shl**{\dt*}\advance\n*\*one
\shl**{\dd*}\advance\n*\*two
\dd*-\dd*\ifnum\n*<\N*\repeat\n*\N*\shl**{0pt}}}

\def\wavebox#1{\setbox0\hbox{#1}\relax
\a*=\wd0\advance\a*14pt\b*=\ht0\advance\b*\dp0\advance\b*14pt\relax
\hbox{\kern9pt\relax
\rmov*(0pt,\ht0){\rmov*(-7pt,7pt){\wlin*\a*(1,0)[+]\wlin*\b*(0,-1)[-]}}\relax
\rmov*(\wd0,-\dp0){\rmov*(7pt,-7pt){\wlin*\a*(-1,0)[+]\wlin*\b*(0,1)[-]}}\relax
\box0\kern9pt}}

\def\rectangle#1(#2,#3){\relax
\lin#1(#2,0)\lin#1(0,#3)\mov#1(0,#3){\lin#1(#2,0)}\mov#1(#2,0){\lin#1(0,#3)}}

\def\dashrectangle#1(#2,#3){\dashlin#1(#2,0)\dashlin#1(0,#3)\relax
\mov#1(0,#3){\dashlin#1(#2,0)}\mov#1(#2,0){\dashlin#1(0,#3)}}

\def\waverectangle#1(#2,#3){\L*=#1\Lengthunit\a*=#2\L*\b*=#3\L*
\ifdim\a*<0pt\a*-\a*\def\x*{-1}\else\def\x*{1}\fi
\ifdim\b*<0pt\b*-\b*\def\y*{-1}\else\def\y*{1}\fi
\wlin*\a*(\x*,0)[-]\wlin*\b*(0,\y*)[+]\relax
\mov#1(0,#3){\wlin*\a*(\x*,0)[+]}\mov#1(#2,0){\wlin*\b*(0,\y*)[-]}}

\def\calcparab*{\ifnum\n*>\m*\k*\N*\advance\k*-\n*\else\k*\n*\fi
\a*=\the\k* sp\a*=10\a*\b*\dm*\advance\b*-\a*\k*\b*
\a*=\the\*ths\b*\divide\a*\l*\multiply\a*\k*
\divide\a*\l*\k*\*ths\r*\a*\advance\k*-\r*\dt*=\the\k*\L*}

\def\arcto#1(#2,#3)[#4]{\rlap{\toks0={#2}\toks1={#3}\calcnum*#1(#2,#3)\relax
\dm*=135sp\dm*=#1\dm*\d**=#1\Lengthunit\ifdim\dm*<0pt\dm*-\dm*\fi
\multiply\dm*\r*\a*=.3\dm*\a*=#4\a*\ifdim\a*<0pt\a*-\a*\fi
\advance\dm*\a*\N*\dm*\divide\N*10000\relax
\divide\N*2\multiply\N*2\advance\N*\*one
\L*=-.25\d**\L*=#4\L*\divide\d**\N*\divide\L*\*ths
\m*\N*\divide\m*2\dm*=\the\m*5sp\l*\dm*\sm*\n*\*one\loop
\calcparab*\shl**{-\dt*}\advance\n*1\ifnum\n*<\N*\repeat}}

\def\arrarcto#1(#2,#3)[#4]{\L*=#1\Lengthunit\L*=.54\L*
\arcto#1(#2,#3)[#4]\rmov*(#2\L*,#3\L*){\d*=.457\L*\d*=#4\d*\d**-\d*
\rmov*(#3\d**,#2\d*){\arrow.02(#2,#3)}}}

\def\dasharcto#1(#2,#3)[#4]{\rlap{\toks0={#2}\toks1={#3}\relax
\calcnum*#1(#2,#3)\dm*=\the\N*5sp\a*=.3\dm*\a*=#4\a*\ifdim\a*<0pt\a*-\a*\fi
\advance\dm*\a*\N*\dm*
\divide\N*20\multiply\N*2\advance\N*1\d**=#1\Lengthunit
\L*=-.25\d**\L*=#4\L*\divide\d**\N*\divide\L*\*ths
\m*\N*\divide\m*2\dm*=\the\m*5sp\l*\dm*
\sm*\n*\*one\loop\calcparab*
\shl**{-\dt*}\advance\n*1\ifnum\n*>\N*\else\calcparab*
\sh*(#2,#3){\xL*=#3\dt* \yL*=#2\dt*
\rx* \the\cos*\xL* \tmp* \the\sin*\yL* \advance\rx*\tmp*
\ry* \the\cos*\yL* \tmp* \the\sin*\xL* \advance\ry*-\tmp*
\kern\rx*\lower\ry*\hbox{\sm*}}\fi
\advance\n*1\ifnum\n*<\N*\repeat}}

\def\*shl*#1{\c*=\the\n*\d**\advance\c*#1\a**\d*\dt*\advance\d*#1\b**
\a*=\the\toks0\c*\b*=\the\toks1\d*\advance\a*-\b*
\b*=\the\toks1\c*\d*=\the\toks0\d*\advance\b*\d*
\rx* \the\cos*\a* \tmp* \the\sin*\b* \advance\rx*-\tmp*
\ry* \the\cos*\b* \tmp* \the\sin*\a* \advance\ry*\tmp*
\raise\ry*\rlap{\kern\rx*\unhcopy\spl*}}

\def\calcnormal*#1{\b**=10000sp\a**\b**\k*\n*\advance\k*-\m*
\multiply\a**\k*\divide\a**\m*\a**=#1\a**\ifdim\a**<0pt\a**-\a**\fi
\ifdim\a**>\b**\d*=.96\a**\advance\d*.4\b**
\else\d*=.96\b**\advance\d*.4\a**\fi
\d*=.01\d*\r*\d*\divide\a**\r*\divide\b**\r*
\ifnum\k*<0\a**-\a**\fi\d*=#1\d*\ifdim\d*<0pt\b**-\b**\fi
\k*\a**\a**=\the\k*\dd*\k*\b**\b**=\the\k*\dd*}

\def\wavearcto#1(#2,#3)[#4]{\rlap{\toks0={#2}\toks1={#3}\relax
\calcnum*#1(#2,#3)\c*=\the\N*5sp\a*=.4\c*\a*=#4\a*\ifdim\a*<0pt\a*-\a*\fi
\advance\c*\a*\N*\c*\divide\N*20\multiply\N*2\advance\N*-1\multiply\N*4\relax
\d**=#1\Lengthunit\dd*=.012\d**
\divide\dd*\*ths \multiply\dd*\magnitude
\ifdim\d**<0pt\d**-\d**\fi\L*=.25\d**
\divide\d**\N*\divide\dd*\N*\L*=#4\L*\divide\L*\*ths
\m*\N*\divide\m*2\dm*=\the\m*0sp\l*\dm*
\sm*\n*\*one\loop\calcnormal*{#4}\calcparab*
\*shl*{1}\advance\n*\*one\calcparab*
\*shl*{1.3}\advance\n*\*one\calcparab*
\*shl*{1}\advance\n*2\dd*-\dd*\ifnum\n*<\N*\repeat\n*\N*\shl**{0pt}}}

\def\triangarcto#1(#2,#3)[#4]{\rlap{\toks0={#2}\toks1={#3}\relax
\calcnum*#1(#2,#3)\c*=\the\N*5sp\a*=.4\c*\a*=#4\a*\ifdim\a*<0pt\a*-\a*\fi
\advance\c*\a*\N*\c*\divide\N*20\multiply\N*2\advance\N*-1\multiply\N*2\relax
\d**=#1\Lengthunit\dd*=.012\d**
\divide\dd*\*ths \multiply\dd*\magnitude
\ifdim\d**<0pt\d**-\d**\fi\L*=.25\d**
\divide\d**\N*\divide\dd*\N*\L*=#4\L*\divide\L*\*ths
\m*\N*\divide\m*2\dm*=\the\m*0sp\l*\dm*
\sm*\n*\*one\loop\calcnormal*{#4}\calcparab*
\*shl*{1}\advance\n*2\dd*-\dd*\ifnum\n*<\N*\repeat\n*\N*\shl**{0pt}}}

\def\hr*#1{\L*=\xscale\Lengthunit\ifnum
\angle**=0\clap{\vrule width#1\L* height.1pt}\else
\L*=#1\L*\L*=.5\L*\rmov*(-\L*,0pt){\sm*}\rmov*(\L*,0pt){\sl*}\fi}

\def\shade#1[#2]{\rlap{\Lengthunit=#1\Lengthunit
\special{em:linewidth .001pt}\relax
\mov(0,#2.05){\hr*{.994}}\mov(0,#2.1){\hr*{.980}}\relax
\mov(0,#2.15){\hr*{.953}}\mov(0,#2.2){\hr*{.916}}\relax
\mov(0,#2.25){\hr*{.867}}\mov(0,#2.3){\hr*{.798}}\relax
\mov(0,#2.35){\hr*{.715}}\mov(0,#2.4){\hr*{.603}}\relax
\mov(0,#2.45){\hr*{.435}}\special{em:linewidth \the\linwid*}}}

\def\dshade#1[#2]{\rlap{\special{em:linewidth .001pt}\relax
\Lengthunit=#1\Lengthunit\if#2-\def\t*{+}\else\def\t*{-}\fi
\mov(0,\t*.025){\relax
\mov(0,#2.05){\hr*{.995}}\mov(0,#2.1){\hr*{.988}}\relax
\mov(0,#2.15){\hr*{.969}}\mov(0,#2.2){\hr*{.937}}\relax
\mov(0,#2.25){\hr*{.893}}\mov(0,#2.3){\hr*{.836}}\relax
\mov(0,#2.35){\hr*{.760}}\mov(0,#2.4){\hr*{.662}}\relax
\mov(0,#2.45){\hr*{.531}}\mov(0,#2.5){\hr*{.320}}\relax
\special{em:linewidth \the\linwid*}}}}

\def\vdot{\rlap{\kern-1.9pt\lower1.8pt\hbox{$\scriptstyle\bullet$}}}
\def\vtimes{\rlap{\kern-3pt\lower1.8pt\hbox{$\scriptstyle\times$}}}
\def\vDot{\rlap{\kern-2.3pt\lower2.7pt\hbox{$\bullet$}}}
\def\vTimes{\rlap{\kern-3.6pt\lower2.4pt\hbox{$\times$}}}

\def\arc(#1)[#2,#3]{{\k*=#2\l*=#3\m*=\l*
\advance\m*-6\ifnum\k*>\l*\relax\else
{\rotate(#2)\mov(#1,0){\sm*}}\loop
\ifnum\k*<\m*\advance\k*5{\rotate(\k*)\mov(#1,0){\sl*}}\repeat
{\rotate(#3)\mov(#1,0){\sl*}}\fi}}

\def\dasharc(#1)[#2,#3]{{\k**=#2\n*=#3\advance\n*-1\advance\n*-\k**
\L*=1000sp\L*#1\L* \multiply\L*\n* \multiply\L*\Nhalfperiods
\divide\L*57\N*\L* \divide\N*2000\ifnum\N*=0\N*1\fi
\r*\n*  \divide\r*\N* \ifnum\r*<2\r*2\fi
\m**\r* \divide\m**2 \l**\r* \advance\l**-\m** \N*\n* \divide\N*\r*
\k**\r* \multiply\k**\N* \dn*\n* \advance\dn*-\k**
\divide\dn*2\advance\dn*\*one
\r*\l** \divide\r*2\advance\dn*\r* \advance\N*-2\k**#2\relax
\ifnum\l**<6{\rotate(#2)\mov(#1,0){\sm*}}\advance\k**\dn*
{\rotate(\k**)\mov(#1,0){\sl*}}\advance\k**\m**
{\rotate(\k**)\mov(#1,0){\sm*}}\loop
\advance\k**\l**{\rotate(\k**)\mov(#1,0){\sl*}}\advance\k**\m**
{\rotate(\k**)\mov(#1,0){\sm*}}\advance\N*-1\ifnum\N*>0\repeat
{\rotate(#3)\mov(#1,0){\sl*}}\else\advance\k**\dn*
\arc(#1)[#2,\k**]\loop\advance\k**\m** \r*\k**
\advance\k**\l** {\arc(#1)[\r*,\k**]}\relax
\advance\N*-1\ifnum\N*>0\repeat
\advance\k**\m**\arc(#1)[\k**,#3]\fi}}

\def\triangarc#1(#2)[#3,#4]{{\k**=#3\n*=#4\advance\n*-\k**
\L*=1000sp\L*#2\L* \multiply\L*\n* \multiply\L*\Nhalfperiods
\divide\L*57\N*\L* \divide\N*1000\ifnum\N*=0\N*1\fi
\d**=#2\Lengthunit \d*\d** \divide\d*57\multiply\d*\n*
\r*\n*  \divide\r*\N* \ifnum\r*<2\r*2\fi
\m**\r* \divide\m**2 \l**\r* \advance\l**-\m** \N*\n* \divide\N*\r*
\dt*\d* \divide\dt*\N* \dt*.5\dt* \dt*#1\dt*
\divide\dt*1000\multiply\dt*\magnitude
\k**\r* \multiply\k**\N* \dn*\n* \advance\dn*-\k** \divide\dn*2\relax
\r*\l** \divide\r*2\advance\dn*\r* \advance\N*-1\k**#3\relax
{\rotate(#3)\mov(#2,0){\sm*}}\advance\k**\dn*
{\rotate(\k**)\mov(#2,0){\sl*}}\advance\k**-\m**\advance\l**\m**\loop\dt*-\dt*
\d*\d** \advance\d*\dt*
\advance\k**\l**{\rotate(\k**)\rmov*(\d*,0pt){\sl*}}%
\advance\N*-1\ifnum\N*>0\repeat\advance\k**\m**
{\rotate(\k**)\mov(#2,0){\sl*}}{\rotate(#4)\mov(#2,0){\sl*}}}}

\def\wavearc#1(#2)[#3,#4]{{\k**=#3\n*=#4\advance\n*-\k**
\L*=4000sp\L*#2\L* \multiply\L*\n* \multiply\L*\Nhalfperiods
\divide\L*57\N*\L* \divide\N*1000\ifnum\N*=0\N*1\fi
\d**=#2\Lengthunit \d*\d** \divide\d*57\multiply\d*\n*
\r*\n*  \divide\r*\N* \ifnum\r*=0\r*1\fi
\m**\r* \divide\m**2 \l**\r* \advance\l**-\m** \N*\n* \divide\N*\r*
\dt*\d* \divide\dt*\N* \dt*.7\dt* \dt*#1\dt*
\divide\dt*1000\multiply\dt*\magnitude
\k**\r* \multiply\k**\N* \dn*\n* \advance\dn*-\k** \divide\dn*2\relax
\divide\N*4\advance\N*-1\k**#3\relax
{\rotate(#3)\mov(#2,0){\sm*}}\advance\k**\dn*
{\rotate(\k**)\mov(#2,0){\sl*}}\advance\k**-\m**\advance\l**\m**\loop\dt*-\dt*
\d*\d** \advance\d*\dt* \dd*\d** \advance\dd*1.3\dt*
\advance\k**\r*{\rotate(\k**)\rmov*(\d*,0pt){\sl*}}\relax
\advance\k**\r*{\rotate(\k**)\rmov*(\dd*,0pt){\sl*}}\relax
\advance\k**\r*{\rotate(\k**)\rmov*(\d*,0pt){\sl*}}\relax
\advance\k**\r*
\advance\N*-1\ifnum\N*>0\repeat\advance\k**\m**
{\rotate(\k**)\mov(#2,0){\sl*}}{\rotate(#4)\mov(#2,0){\sl*}}}}

\def\gmov*#1(#2,#3)#4{\rlap{\L*=#1\Lengthunit
\xL*=#2\L* \yL*=#3\L*
\rx* \gcos*\xL* \tmp* \gsin*\yL* \advance\rx*-\tmp*
\ry* \gcos*\yL* \tmp* \gsin*\xL* \advance\ry*\tmp*
\rx*=\xscale\rx* \ry*=\yscale\ry*
\xL* \the\cos*\rx* \tmp* \the\sin*\ry* \advance\xL*-\tmp*
\yL* \the\cos*\ry* \tmp* \the\sin*\rx* \advance\yL*\tmp*
\kern\xL*\raise\yL*\hbox{#4}}}

\def\rgmov*(#1,#2)#3{\rlap{\xL*#1\yL*#2\relax
\rx* \gcos*\xL* \tmp* \gsin*\yL* \advance\rx*-\tmp*
\ry* \gcos*\yL* \tmp* \gsin*\xL* \advance\ry*\tmp*
\rx*=\xscale\rx* \ry*=\yscale\ry*
\xL* \the\cos*\rx* \tmp* \the\sin*\ry* \advance\xL*-\tmp*
\yL* \the\cos*\ry* \tmp* \the\sin*\rx* \advance\yL*\tmp*
\kern\xL*\raise\yL*\hbox{#3}}}

\def\Earc(#1)[#2,#3][#4,#5]{{\k*=#2\l*=#3\m*=\l*
\advance\m*-6\ifnum\k*>\l*\relax\else\def\xscale{#4}\def\yscale{#5}\relax
{\angle**0\rotate(#2)}\gmov*(#1,0){\sm*}\loop
\ifnum\k*<\m*\advance\k*5\relax
{\angle**0\rotate(\k*)}\gmov*(#1,0){\sl*}\repeat
{\angle**0\rotate(#3)}\gmov*(#1,0){\sl*}\relax
\def\xscale{1}\def\yscale{1}\fi}}

\def\dashEarc(#1)[#2,#3][#4,#5]{{\k**=#2\n*=#3\advance\n*-1\advance\n*-\k**
\L*=1000sp\L*#1\L* \multiply\L*\n* \multiply\L*\Nhalfperiods
\divide\L*57\N*\L* \divide\N*2000\ifnum\N*=0\N*1\fi
\r*\n*  \divide\r*\N* \ifnum\r*<2\r*2\fi
\m**\r* \divide\m**2 \l**\r* \advance\l**-\m** \N*\n* \divide\N*\r*
\k**\r*\multiply\k**\N* \dn*\n* \advance\dn*-\k** \divide\dn*2\advance\dn*\*one
\r*\l** \divide\r*2\advance\dn*\r* \advance\N*-2\k**#2\relax
\ifnum\l**<6\def\xscale{#4}\def\yscale{#5}\relax
{\angle**0\rotate(#2)}\gmov*(#1,0){\sm*}\advance\k**\dn*
{\angle**0\rotate(\k**)}\gmov*(#1,0){\sl*}\advance\k**\m**
{\angle**0\rotate(\k**)}\gmov*(#1,0){\sm*}\loop
\advance\k**\l**{\angle**0\rotate(\k**)}\gmov*(#1,0){\sl*}\advance\k**\m**
{\angle**0\rotate(\k**)}\gmov*(#1,0){\sm*}\advance\N*-1\ifnum\N*>0\repeat
{\angle**0\rotate(#3)}\gmov*(#1,0){\sl*}\def\xscale{1}\def\yscale{1}\else
\advance\k**\dn* \Earc(#1)[#2,\k**][#4,#5]\loop\advance\k**\m** \r*\k**
\advance\k**\l** {\Earc(#1)[\r*,\k**][#4,#5]}\relax
\advance\N*-1\ifnum\N*>0\repeat
\advance\k**\m**\Earc(#1)[\k**,#3][#4,#5]\fi}}

\def\triangEarc#1(#2)[#3,#4][#5,#6]{{\k**=#3\n*=#4\advance\n*-\k**
\L*=1000sp\L*#2\L* \multiply\L*\n* \multiply\L*\Nhalfperiods
\divide\L*57\N*\L* \divide\N*1000\ifnum\N*=0\N*1\fi
\d**=#2\Lengthunit \d*\d** \divide\d*57\multiply\d*\n*
\r*\n*  \divide\r*\N* \ifnum\r*<2\r*2\fi
\m**\r* \divide\m**2 \l**\r* \advance\l**-\m** \N*\n* \divide\N*\r*
\dt*\d* \divide\dt*\N* \dt*.5\dt* \dt*#1\dt*
\divide\dt*1000\multiply\dt*\magnitude
\k**\r* \multiply\k**\N* \dn*\n* \advance\dn*-\k** \divide\dn*2\relax
\r*\l** \divide\r*2\advance\dn*\r* \advance\N*-1\k**#3\relax
\def\xscale{#5}\def\yscale{#6}\relax
{\angle**0\rotate(#3)}\gmov*(#2,0){\sm*}\advance\k**\dn*
{\angle**0\rotate(\k**)}\gmov*(#2,0){\sl*}\advance\k**-\m**
\advance\l**\m**\loop\dt*-\dt* \d*\d** \advance\d*\dt*
\advance\k**\l**{\angle**0\rotate(\k**)}\rgmov*(\d*,0pt){\sl*}\relax
\advance\N*-1\ifnum\N*>0\repeat\advance\k**\m**
{\angle**0\rotate(\k**)}\gmov*(#2,0){\sl*}\relax
{\angle**0\rotate(#4)}\gmov*(#2,0){\sl*}\def\xscale{1}\def\yscale{1}}}

\def\waveEarc#1(#2)[#3,#4][#5,#6]{{\k**=#3\n*=#4\advance\n*-\k**
\L*=4000sp\L*#2\L* \multiply\L*\n* \multiply\L*\Nhalfperiods
\divide\L*57\N*\L* \divide\N*1000\ifnum\N*=0\N*1\fi
\d**=#2\Lengthunit \d*\d** \divide\d*57\multiply\d*\n*
\r*\n*  \divide\r*\N* \ifnum\r*=0\r*1\fi
\m**\r* \divide\m**2 \l**\r* \advance\l**-\m** \N*\n* \divide\N*\r*
\dt*\d* \divide\dt*\N* \dt*.7\dt* \dt*#1\dt*
\divide\dt*1000\multiply\dt*\magnitude
\k**\r* \multiply\k**\N* \dn*\n* \advance\dn*-\k** \divide\dn*2\relax
\divide\N*4\advance\N*-1\k**#3\def\xscale{#5}\def\yscale{#6}\relax
{\angle**0\rotate(#3)}\gmov*(#2,0){\sm*}\advance\k**\dn*
{\angle**0\rotate(\k**)}\gmov*(#2,0){\sl*}\advance\k**-\m**
\advance\l**\m**\loop\dt*-\dt*
\d*\d** \advance\d*\dt* \dd*\d** \advance\dd*1.3\dt*
\advance\k**\r*{\angle**0\rotate(\k**)}\rgmov*(\d*,0pt){\sl*}\relax
\advance\k**\r*{\angle**0\rotate(\k**)}\rgmov*(\dd*,0pt){\sl*}\relax
\advance\k**\r*{\angle**0\rotate(\k**)}\rgmov*(\d*,0pt){\sl*}\relax
\advance\k**\r*
\advance\N*-1\ifnum\N*>0\repeat\advance\k**\m**
{\angle**0\rotate(\k**)}\gmov*(#2,0){\sl*}\relax
{\angle**0\rotate(#4)}\gmov*(#2,0){\sl*}\def\xscale{1}\def\yscale{1}}}

\newcount\CatcodeOfAtSign
\CatcodeOfAtSign=\the\catcode`\@
\catcode`\@=11
\def\@arc#1[#2][#3]{\rlap{\Lengthunit=#1\Lengthunit
\sm*\l*arc(#2.1914,#3.0381)[#2][#3]\relax
\mov(#2.1914,#3.0381){\l*arc(#2.1622,#3.1084)[#2][#3]}\relax
\mov(#2.3536,#3.1465){\l*arc(#2.1084,#3.1622)[#2][#3]}\relax
\mov(#2.4619,#3.3086){\l*arc(#2.0381,#3.1914)[#2][#3]}}}

\def\dash@arc#1[#2][#3]{\rlap{\Lengthunit=#1\Lengthunit
\d*arc(#2.1914,#3.0381)[#2][#3]\relax
\mov(#2.1914,#3.0381){\d*arc(#2.1622,#3.1084)[#2][#3]}\relax
\mov(#2.3536,#3.1465){\d*arc(#2.1084,#3.1622)[#2][#3]}\relax
\mov(#2.4619,#3.3086){\d*arc(#2.0381,#3.1914)[#2][#3]}}}

\def\wave@arc#1[#2][#3]{\rlap{\Lengthunit=#1\Lengthunit
\w*lin(#2.1914,#3.0381)\relax
\mov(#2.1914,#3.0381){\w*lin(#2.1622,#3.1084)}\relax
\mov(#2.3536,#3.1465){\w*lin(#2.1084,#3.1622)}\relax
\mov(#2.4619,#3.3086){\w*lin(#2.0381,#3.1914)}}}

\def\bezier#1(#2,#3)(#4,#5)(#6,#7){\N*#1\l*\N* \advance\l*\*one
\d* #4\Lengthunit \advance\d* -#2\Lengthunit \multiply\d* \*two
\b* #6\Lengthunit \advance\b* -#2\Lengthunit
\advance\b*-\d* \divide\b*\N*
\d** #5\Lengthunit \advance\d** -#3\Lengthunit \multiply\d** \*two
\b** #7\Lengthunit \advance\b** -#3\Lengthunit
\advance\b** -\d** \divide\b**\N*
\mov(#2,#3){\sm*{\loop\ifnum\m*<\l*
\a*\m*\b* \advance\a*\d* \divide\a*\N* \multiply\a*\m*
\a**\m*\b** \advance\a**\d** \divide\a**\N* \multiply\a**\m*
\rmov*(\a*,\a**){\unhcopy\spl*}\advance\m*\*one\repeat}}}

\catcode`\*=12

\newcount\n@ast
\def\n@ast@#1{\n@ast0\relax\get@ast@#1\end}
\def\get@ast@#1{\ifx#1\end\let\next\relax\else
\ifx#1*\advance\n@ast1\fi\let\next\get@ast@\fi\next}

\newif\if@up \newif\if@dwn
\def\up@down@#1{\@upfalse\@dwnfalse
\if#1u\@uptrue\fi\if#1U\@uptrue\fi\if#1+\@uptrue\fi
\if#1d\@dwntrue\fi\if#1D\@dwntrue\fi\if#1-\@dwntrue\fi}

\def\halfcirc#1(#2)[#3]{{\Lengthunit=#2\Lengthunit\up@down@{#3}\relax
\if@up\mov(0,.5){\@arc[-][-]\@arc[+][-]}\fi
\if@dwn\mov(0,-.5){\@arc[-][+]\@arc[+][+]}\fi
\def\lft{\mov(0,.5){\@arc[-][-]}\mov(0,-.5){\@arc[-][+]}}\relax
\def\rght{\mov(0,.5){\@arc[+][-]}\mov(0,-.5){\@arc[+][+]}}\relax
\if#3l\lft\fi\if#3L\lft\fi\if#3r\rght\fi\if#3R\rght\fi
\n@ast@{#1}\relax
\ifnum\n@ast>0\if@up\shade[+]\fi\if@dwn\shade[-]\fi\fi
\ifnum\n@ast>1\if@up\dshade[+]\fi\if@dwn\dshade[-]\fi\fi}}

\def\halfdashcirc(#1)[#2]{{\Lengthunit=#1\Lengthunit\up@down@{#2}\relax
\if@up\mov(0,.5){\dash@arc[-][-]\dash@arc[+][-]}\fi
\if@dwn\mov(0,-.5){\dash@arc[-][+]\dash@arc[+][+]}\fi
\def\lft{\mov(0,.5){\dash@arc[-][-]}\mov(0,-.5){\dash@arc[-][+]}}\relax
\def\rght{\mov(0,.5){\dash@arc[+][-]}\mov(0,-.5){\dash@arc[+][+]}}\relax
\if#2l\lft\fi\if#2L\lft\fi\if#2r\rght\fi\if#2R\rght\fi}}

\def\halfwavecirc(#1)[#2]{{\Lengthunit=#1\Lengthunit\up@down@{#2}\relax
\if@up\mov(0,.5){\wave@arc[-][-]\wave@arc[+][-]}\fi
\if@dwn\mov(0,-.5){\wave@arc[-][+]\wave@arc[+][+]}\fi
\def\lft{\mov(0,.5){\wave@arc[-][-]}\mov(0,-.5){\wave@arc[-][+]}}\relax
\def\rght{\mov(0,.5){\wave@arc[+][-]}\mov(0,-.5){\wave@arc[+][+]}}\relax
\if#2l\lft\fi\if#2L\lft\fi\if#2r\rght\fi\if#2R\rght\fi}}

\catcode`\*=11

\def\Circle#1(#2){\halfcirc#1(#2)[u]\halfcirc#1(#2)[d]\n@ast@{#1}\relax
\ifnum\n@ast>0\L*=\xscale\Lengthunit
\ifnum\angle**=0\clap{\vrule width#2\L* height.1pt}\else
\L*=#2\L*\L*=.5\L*\special{em:linewidth .001pt}\relax
\rmov*(-\L*,0pt){\sm*}\rmov*(\L*,0pt){\sl*}\relax
\special{em:linewidth \the\linwid*}\fi\fi}

\catcode`\*=12

\def\wavecirc(#1){\halfwavecirc(#1)[u]\halfwavecirc(#1)[d]}

\def\dashcirc(#1){\halfdashcirc(#1)[u]\halfdashcirc(#1)[d]}

\def\xscale{1}
\def\yscale{1}

\def\Ellipse#1(#2)[#3,#4]{\def\xscale{#3}\def\yscale{#4}\relax
\Circle#1(#2)\def\xscale{1}\def\yscale{1}}

\def\dashEllipse(#1)[#2,#3]{\def\xscale{#2}\def\yscale{#3}\relax
\dashcirc(#1)\def\xscale{1}\def\yscale{1}}

\def\waveEllipse(#1)[#2,#3]{\def\xscale{#2}\def\yscale{#3}\relax
\wavecirc(#1)\def\xscale{1}\def\yscale{1}}

\def\halfEllipse#1(#2)[#3][#4,#5]{\def\xscale{#4}\def\yscale{#5}\relax
\halfcirc#1(#2)[#3]\def\xscale{1}\def\yscale{1}}

\def\halfdashEllipse(#1)[#2][#3,#4]{\def\xscale{#3}\def\yscale{#4}\relax
\halfdashcirc(#1)[#2]\def\xscale{1}\def\yscale{1}}

\def\halfwaveEllipse(#1)[#2][#3,#4]{\def\xscale{#3}\def\yscale{#4}\relax
\halfwavecirc(#1)[#2]\def\xscale{1}\def\yscale{1}}

\catcode`\@=\the\CatcodeOfAtSign

\section{Introduction}

Different processes of the production and decay of heavy mesons
consisting of heavy $b$ and $c$ quarks provide the means for
revealing the role of the color and spin quark forces. The aim of
many present experiments consists in the increase of experimental
accuracy what is important for the detailed comparison of
different existing theoretical approaches to the heavy quark
problems \cite{BFY,UFN1,QWG,Bali}. The exclusive production of a pair of
doubly heavy mesons with $c$-quarks in $e^+e^-$ annihilation has
attracted considerable attention in the last years. This is due
to the fact that the cross section of the process $e^++e^-\to
J/\Psi+\eta_c$ which was measured in the experiments on Babar and
Belle detectors at the energy $\sqrt{s}=10.6$ GeV
\begin{equation}
\sigma(e^+e^-\to J/\Psi+\eta_c)\times{\cal B}(\eta_c\to\geq
2~charged)= \Biggl\{{{25.6\pm 2.8\pm
3.4~~~[5,6]}\atop{17.6\pm
2.8^{+1.5}_{-2.1}~~~[7]}}
\end{equation}
leads to a discrepancy with the theoretical calculation in the framework of
nonrelativistic QCD (NRQCD) by an order of magnitude
\cite{Belle1,Belle, BaBar,BL1,Chao,Chao1}. This conclusion is based on calculations
in which the relative momenta of heavy quarks and bound state effects
in the production amplitude
were not taken into account. A set of calculations was performed to
improve the nonrelativistic approximation for the process. In particular, relativistic corrections to the cross section
$\sigma(e^+e^-\to J/\Psi\eta_c)$ were considered in a color singlet
model in Ref.\cite{BL1} using the methods of NRQCD \cite{BBL}. It
was obtained here that the relativistic corrections increase the cross
section by a factor 2.4 for the production $J/\Psi+\eta_c$.
Another attempt to take into account the relativistic corrections
was done in the framework of the light-cone formalism \cite{Ma,BC,BLL}.
Here it was shown that the discrepancy between the experiment and
the theory can be eliminated completely by considering the intrinsic
motion of heavy quarks forming the doubly heavy mesons. Thereupon,
perturbative corrections of order $\alpha_s$ to the production
amplitude were calculated in Ref.\cite{ZGC} increasing the cross
section by a factor 1.8. On account of different values of
relativistic corrections obtained in Refs.\cite{BL1,Ma,BC,BLL} and
the importance of a relativistic consideration of the process
$e^++e^-\to J/\Psi+\eta_c$ in solving the doubly heavy meson
production problem, we have performed a new investigation of
relativistic and bound state effects. As in our recent papers 
\cite{rqm4,apm2005}, this investigation is based on the relativistic quark model which provides the solution in many
tasks of heavy quark physics.
In the above quoted papers Refs.\cite{rqm4,apm2005}
we have demonstrated how the original amplitude, describing the
physical process, must be transformed in order to preserve the
relativistic plus bound state corrections connected with the
one-particle wave functions and the wave function of a two-particle
bound state. In the present paper we shall extend the method of 
Refs.\cite{rqm4,apm2005} to the case of the production of a pair 
(${\cal P}+{\cal V}$)  of doubly heavy mesons containing quarks of different 
flavours $b$ and $c$. In particular, we will consider the internal 
motion of heavy
quarks in both produced pseudoscalar ${\cal P}$ and
vector ${\cal V}$ mesons. The paper is organized as follows: In Sec.II we present the general formalism and the
basic relations of our method which are required in order
to formulate the relativistic amplitude
for the production of doubly heavy mesons. In Sec.III we derive the
analytical expressions for the corresponding cross sections
and make numerical estimations exploiting the relativistic quark model.
Conclusions and discussion of the results are given in Sec.IV.
The construction of the quasipotential
heavy quark distribution amplitude in the light-front variables
is included in the Appendix.

\section{General formalism}

The production of a pair of doubly heavy mesons in a color singlet
model from $e^+e^-$ annihilation contains two stages after the
transition of the virtual photon $\gamma^\ast$ into a quark-antiquark
pair $(Q_1\bar Q_1)$. In the first stage examined on the basis
of perturbative quantum chromodynamics (QCD),
one of the heavy quarks
($Q_1$ or $\bar Q_1$) emits a gluon with sufficiently large
energy $\sim\sqrt{s}$ which then transforms to another heavy
quark-antiquark pair $(Q_2\bar Q_2)$.
These four quarks can combine with a definite probability
into a pair of $S$ - wave
pseudoscalar $(\bar Q_1Q_2)_{S=0}$ and vector $(Q_1\bar
Q_2)_{S=1}$ doubly heavy mesons. The second 
nonperturbative stage of this process
involves the formation of heavy quark bound states from heavy
quarks. In the quasipotential approach to the relativistic quark
model we can express the invariant transition amplitude for the
described process as a simple convolution of a perturbative
production amplitude of four heavy quarks ${\cal T}(p_1,p_2;q_1,q_2)$,
projected onto the positive energy states, 
and the quasipotential wave functions of a vector meson $\Psi_{\cal
V}(p,P)$ and a pseudoscalar meson $\Psi_{\cal P}(q,Q)$
\cite{savrin}:
\begin{equation}
{\cal M}(p_{-},p_+,P,Q)=\int\frac{d{\bf p}}{(2\pi)^3}\bar\Psi_{\cal
V}(p,P)\int\frac{d{\bf q}}{(2\pi)^3}\bar\Psi_{\cal P}(q,Q)\cdot
{\cal T}_\beta(p_1,p_2;q_1,q_2)\bar v(p_+)\gamma^\alpha
u(p_-)\frac{4\pi\alpha g^{\alpha\beta}} {r^2+i\epsilon},
\end{equation}
where $p_-$, $p_+$ are four momenta of the electron and positron,
$r^2$ = $s$ = $(p_-+p_+)^2$, $p_1$, $p_2$ are four-momenta of $Q_1$ and
$\bar Q_2$ quarks forming the vector doubly heavy meson ${\cal
V}$; $q_1$, $q_2$ are four momenta of $\bar Q_1$ and $Q_2$ forming
the pseudoscalar doubly heavy meson ${\cal P}$. They are defined
in terms of total momenta $P(Q)$ and relative momenta $p(q)$ as
follows:
\begin{equation}
p_{1,2}=\eta_{1,2}P\pm p,~~~(p\cdot
P)=0,~~~\eta_{1,2}=\frac{E_{1,2}^{\cal V}}{M_{\cal V}}=
\frac{M_{\cal V}^2-m_{2,1}^2+m_{1,2}^2}{2M_{\cal V}^2},
\end{equation}
\begin{equation}
q_{1,2}=\rho_{1,2}Q\pm q,~~~(q\cdot
Q)=0,~~~\rho_{1,2}=\frac{E_{1,2}^{\cal P}}{M_{\cal P}}=
\frac{M_{\cal P}^2-m_{2,1}^2+m_{1,2}^2}{2M_{\cal P}^2},
\end{equation}
where $M_{\cal V}=m_1+m_2+W_{\cal V}$, $M_{\cal
P}=m_1+m_2+W_{\cal P}$ are the masses of vector and pseudoscalar
mesons consisting of heavy quarks.

Different color-spin nonperturbative factors entering the
amplitude ${\cal M}(p_-,p_+,P,Q)$ control the production of
doubly heavy mesons. In this process the gluon virtuality is large
$k^2\gg\Lambda_{QCD}^2$, and the strong interaction constant is small
$\alpha_s\ll 1$. There exist four production amplitudes in the
leading order over $\alpha_s$ as explained below in Fig.1. In a color singlet
model the first amplitude ${\cal T}_{1,\beta}(p_1,p_2;q_1,q_2)$ takes the
form:
\begin{equation}
{\cal T}_{1,\beta}(p_1,p_2;q_1,q_2)=\frac{16\pi\alpha_s}{3}\bar
u_1(p_1)\gamma_\mu\frac{(\hat r-\hat q_1+m_1)}
{(r-q_1)^2-m_1^2+i\epsilon}\gamma_\beta v_1(q_1)\bar
u_2(q_2)\gamma_\nu v_2(p_2) D_{\mu\nu}(k).
\end{equation}
The color factor
$\delta_{il}\delta_{kj}(T^a)_{ij}(T^a)_{kl}/3=4/3$ is already
extracted in the amplitude (5). To calculate relativistic effects
we have to keep all factors in the amplitude (2) with the
relative motion momenta $p$ and $q$. We have to take into account
also the bound state corrections which are determined by the
binding energies $W_{\cal V}$ and $W_{\cal P}$.

\begin{figure}[htbp]
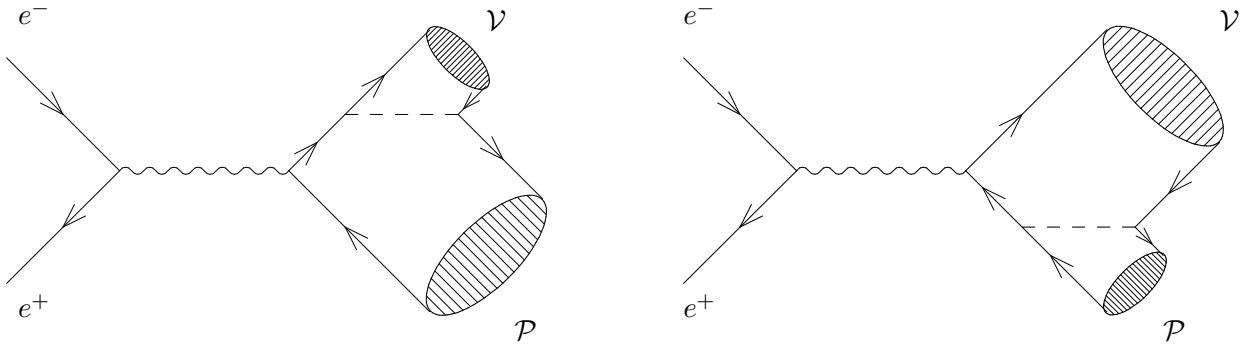

\magnitude=1000 \GRAPH(hsize=15){
\mov(0.,-1){\lin(1,1)}%
\mov(0.5,0.5){\lin(-0.1,0.2)}%
\mov(0.5,0.5){\lin(-0.2,0.1)}%
\mov(0.5,-0.5){\lin(0.1,0.2)}%
\mov(0.5,-0.5){\lin(0.2,0.1)}%
\mov(0,1.){\lin(1,-1)}%
\mov(1,0){\wavelin(1.5,0)}%
\mov(2.75,0.25){\lin(-0.2,-0.1)}%
\mov(2.75,0.25){\lin(-0.1,-0.2)}%
\mov(3.35,0.85){\lin(-0.1,-0.2)}%
\mov(3.35,0.85){\lin(-0.2,-0.1)}%
\mov(3.,-0.5){\lin(0.2,-0.1)}%
\mov(3.,-0.5){\lin(0.1,-0.2)}%
\mov(2.5,0){\lin(1.25,1.25)}%
\mov(4.4,0.1){\lin(-0.2,0.1)}%
\mov(4.4,0.1){\lin(-0.1,0.2)}%
\mov(10.3,-0.2){\lin(0.2,0.1)}%
\mov(10.3,-0.2){\lin(0.1,0.2)}%
\mov(10.15,-0.65){\lin(-0.14,0.07)}%
\mov(10.15,-0.65){\lin(-0.07,0.14)}%
\mov(4.05,0.55){\lin(0.14,0.07)}%
\mov(4.05,0.55){\lin(0.07,0.14)}%
\mov(8.65,-0.15){\lin(0.2,-0.1)}%
\mov(8.65,-0.15){\lin(0.1,-0.2)}%
\mov(9.25,-0.75){\lin(0.2,-0.1)}%
\mov(9.25,-0.75){\lin(0.1,-0.2)}%
\mov(9.,0.5){\lin(-0.2,-0.1)}%
\mov(9,0.5){\lin(-0.1,-0.2)}%
\mov(2.5,0){\lin(1.25,-1.25)}%
\mov(3.,0.5){\dashlin(1.,0)}%
\mov(4.,0.5){\lin(0.25,0.25)}%
\mov(4.,0.5){\lin(0.75,-0.75)}%
\mov(4.,1.){\rotate(45)\Ellipse*(0.73)[0.4,1]}%
\mov(4.25,-0.75){\rotate(-45)\Ellipse*(1.4)[0.4,1]}%
\mov(6.,-1){\lin(1,1)}%
\mov(4.25,1.25){${\cal V}$}%
\mov(4.5,-1.5){${\cal P}$}%
\mov(10.25,-1.5){${\cal P}$}%
\mov(10.75,1.25){${\cal V}$}%
\mov(0.1,1.3){$e^-$}%
\mov(0.1,-1.3){$e^+$}%
\mov(6.,1.3){$e^-$}%
\mov(6.,-1.3){$e^+$}%
\mov(6.5,0.5){\lin(-0.1,0.2)}%
\mov(6.5,0.5){\lin(-0.2,0.1)}%
\mov(6.5,-0.5){\lin(0.1,0.2)}%
\mov(6.5,-0.5){\lin(0.2,0.1)}%
\mov(6,1.){\lin(1,-1)}%
\mov(7,0){\wavelin(1.5,0)}%
\mov(8.5,0){\lin(1.25,1.25)}%
\mov(8.5,0){\lin(1.25,-1.25)}%
\mov(9.,-0.5){\dashlin(1.,0)}%
\mov(10.,-0.5){\lin(0.75,0.75)}%
\mov(10.,-0.5){\lin(0.25,-0.25)}%
\mov(10.,-1.){\rotate(-45)\Ellipse*(0.73)[0.4,1]}%
\mov(10.25,0.75){\rotate(45)\Ellipse*(1.4)[0.4,1]}%
} \vspace{3mm}
\caption{The Feynman diagrams for the production
of a pair of doubly heavy mesons $({\cal P}+{\cal V})$ in
$e^+e^-$ annihilation. The wave line corresponds to the photon
and the dashed line corresponds to the gluon. Two other diagrams can
be obtained by permutations.}
\end{figure}

Clearly, in order to calculate the matrix element (2), one needs
explicit expressions
for bound state wave functions. We construct such wave functions
by solving a relativistic quasipotential equation to the desired
accuracy in the center-of-mass (CM) frame. But in the matrix
element (2) the final meson states have different total momenta
$P$ and $Q$. So, it is necessary to know how to transform the CM
wave function to an arbitrary reference frame. The transformation
law of the bound state wave function from the rest frame to the
moving one with four-momentum $P$ was derived in the
Bethe-Salpeter approach in Ref.\cite{Brodsky} and in the
quasipotential method in Ref.\cite{F1973}. We use the last one
and write the necessary transformation as follows:
\begin{equation}
\Psi_{P}^{\rho\omega}({\bf p})=D_1^{1/2,~\rho\alpha}(R^W_{L_{P}})
D_2^{1/2,~\omega\beta}(R^W_{L_{P}})\Psi_{0}^{\alpha\beta}({\bf p}),
\end{equation}
\begin{displaymath}
\bar\Psi_{P}^{\lambda\sigma}({\bf p})
=\bar\Psi^{\varepsilon\tau}_{0}({\bf p})D_1^{+~1/2,~\varepsilon
\lambda}(R^W_{L_{P}})D_2^{+~1/2,~\tau\sigma}(R^W_{L_{P}}),
\end{displaymath}
where $R^W$ is the Wigner rotation, $L_{P}$ is the Lorentz boost
from the meson rest frame to a moving one, and
the rotation matrix $D^{1/2}(R)$ is defined by
\begin{equation}
{1 \ \ \,0\choose 0 \ \ \,1}D^{1/2}_{1,2}(R^W_{L_{P}})=
S^{-1}({\bf p}_{1,2})S({\bf P})S({\bf p}),
\end{equation}
where the explicit form for the Lorentz transformation matrix of the four-spinor
is
\begin{equation}
S({\bf p})=\sqrt{\frac{\epsilon(p)+m}{2m}}\left(1+\frac{(\bm{\alpha}
{\bf p})} {\epsilon(p)+m}\right).
\end{equation}
For further transformation of the amplitude (2) the following relations
are applied:
\begin{equation}
S_{\alpha\beta}(\Lambda)u^\lambda_\beta(p)=\sum_{\sigma=\pm 1/2}
u^{\sigma}_\alpha(\Lambda p)D^{1/2}_{\sigma\lambda}(R^W_{\Lambda p}),
\end{equation}
\begin{displaymath}
\bar u^\lambda_\beta(p)S^{-1}_{\beta\alpha}(\Lambda)=\sum_{\sigma=\pm 1/2}
D^{+~1/2}_{\lambda\sigma}(R^W_{\Lambda p})\bar u^\sigma_\alpha(\Lambda p).
\end{displaymath}
Using also the transformation property of the Dirac bispinors to the rest frame
\begin{eqnarray}
\bar u_1({\bf p})=\bar u_1(0)\frac{(\hat
p'_1+m_1)}{\sqrt{2\epsilon_1({\bf p}) (\epsilon_1({\bf
p})+m_1)}},~~p'_1=(\epsilon_1,{\bf p}),\cr\cr v_2(-{\bf
p})=\frac{(\hat p'_2-m_2)}{\sqrt{2\epsilon_2({\bf
p})(\epsilon_2({\bf p})+ m_2)}}v_2(0),~~p'_2=(\epsilon_2,-{\bf p}),
\end{eqnarray}
we can introduce the projection operators $\hat\Pi^{{\cal P},{\cal V}}$
onto the states $(Q_1\bar Q_2)$ in the meson with total spin 0 and 1
as follows:
\begin{equation}
\hat\Pi^{{\cal P},{\cal V}}=[v_2(0)\bar
u_1(0)]_{S=0,1}=\gamma_5(\hat\epsilon^\ast)\frac{1+\gamma^0}
{2\sqrt{2}}.
\end{equation}
As a result the doubly heavy meson production amplitude from $e^+e^-$
annihilation takes the form:
\begin{equation}
{\cal M}_1(p_-,p_+,P,Q)=\frac{8\pi^2\alpha\alpha_s}{3s}\bar
v(p_+)\gamma^\beta u(p_-) \int\frac{d{\bf
p}}{(2\pi)^3}\frac{\bar\Psi_0({\bf p})}{\sqrt{2\epsilon_1(p)
(\epsilon_1(p)+m_1)2\epsilon_2(p)(\epsilon_2(p)+m_2)}}\times
\end{equation}
\begin{displaymath}
\times\int\frac{d{\bf q}}{(2\pi)^3}\frac{\bar\Psi_0({\bf q})}
{\sqrt{2\epsilon_1(q)(\epsilon_1(q)+m_1)2\epsilon_2(q)(\epsilon_2(q)+m_2)}}
Sp\Bigl\{(\hat{\tilde p_2}-m_2)\hat{\tilde\epsilon}^\ast(1+\hat v_1)
(\hat{\tilde p_1}+m_1)\times
\end{displaymath}
\begin{displaymath}
\times\gamma_\mu\frac{(\hat r-\hat q_1+m_1)}{(r-q_1)^2-m_1^2+i\epsilon}
\gamma_\beta(\hat{\tilde q_1}-m_1)\gamma_5(1+\hat v_2)(\hat{\tilde q_2}+m_2)
\gamma_\nu\Bigr\}D_{\mu\nu}(k),
\end{displaymath}
where the four-vectors $\tilde\epsilon$, $\tilde p_{1,2}$,
$\tilde q_{1,2}$ are given by:
\begin{equation}
\tilde\epsilon=L_P(0,{\mathstrut\bm\epsilon})=\left({\mathstrut\bm\epsilon}
{\bf v},{\mathstrut\bm\epsilon}+\frac{({\mathstrut\bm\epsilon}{\bf
v}){\bf v}} {1+v^0}\right),
\end{equation}
\begin{displaymath}
\hat{\tilde p}_{1,2}=S(L_P)\hat p'_{1,2}S^{-1}(L_P),~S(L_P)(1\pm\gamma^0)
S^{-1}(L_P)= \pm(\hat v_1\pm 1),~\hat v_1=\frac{\hat P}{M_{\cal V}},
\end{displaymath}
\begin{displaymath}
\hat{\tilde q}_{1,2}=S(L_Q)\hat
q'_{1,2}S^{-1}(L_Q),~S(L_Q)(1\pm\gamma^0) S^{-1}(L_Q)= \pm(\hat
v_2\pm 1),~\hat v_2=\frac{\hat Q}{M_{\cal P}}.
\end{displaymath}
In order to make optimum use of the expression (12), we rearrange
the bispinor contractions in the numerator of Eq.(12) and thus extracting
in a more evident form the relative momenta $p$ and $q$ of heavy
quarks:
\begin{equation}
{\cal M}_1(p_-,p_+,P,Q)=\frac{8\pi^2\alpha\alpha_s}{3s}\bar
v(p_+)\gamma^\beta u(p_-) \int\frac{d{\bf
p}}{(2\pi)^3}\frac{\bar\Psi_0^{\cal V}({\bf
p})}{\sqrt{\frac{\epsilon_1(p)}{m_1}
\frac{(\epsilon_1(p)+m_1)}{2m_1}\frac{\epsilon_2(p)}{m_2}
\frac{(\epsilon_2(p)+m_2)}{2m_2}}}\times
\end{equation}
\begin{displaymath}
\times\int\frac{d{\bf q}}{(2\pi)^3}\frac{\bar\Psi_0^{\cal P}({\bf
q})}
{\sqrt{\frac{\epsilon_1(q)}{m_1}\frac{(\epsilon_1(q)+m_1)}{2m_1}
\frac{\epsilon_2(q)}{m_2}\frac{(\epsilon_2(q)+m_2)}{2m_2}}}
Sp\Biggl\{\left[\frac{\hat v_1-1}{2}+\hat v_1\frac{{\bf
p}^2}{2m_2(\epsilon_2(p)+ m_2)}-\frac{\hat{p}}{2m_2}\right]\hat{\tilde\epsilon}^\ast(1+\hat v_1)\times
\end{displaymath}
\begin{displaymath}
\times\left[\frac{\hat v_1+1}{2}+\hat v_1\frac{{\bf p}^2}{2m_1(\epsilon_1(p)+
m_1)}+\frac{\hat{p}}{2m_1}\right]\gamma_\mu\frac{(\hat r-\hat q_1+m_1)}
{(r-q_1)^2-m_1^2+i\epsilon}\gamma_\beta D_{\mu\nu}(k)\times
\end{displaymath}
\begin{displaymath}
\times\left[\frac{\hat v_2-1}{2}+\hat v_2\frac{{\bf q}^2}{2m_1(\epsilon_1(q)+
m_1)}+\frac{\hat{q}}{2m_1}\right]\gamma_5(1+\hat v_2)
\left[\frac{\hat v_2+1}{2}+\hat v_2\frac{{\bf q}^2}{2m_2(\epsilon_2(q)+
m_2)}-\frac{\hat{q}}{2m_2}\right]\gamma_\nu\Biggr\}.
\end{displaymath}

The production amplitude (14) contains relativistic corrections of two
species. The first type of corrections are held in the quark interaction
operator. They can be accounted for by means of the numerical solution of the
Schr\"odinger equation with the relevant potential. The second type corrections
are determined by several functions depending on the momenta of relative
motion of quarks ${\bf p}$ and ${\bf q}$ including the gluon propagator
$D_{\mu\nu}(k)$, relativistic bispinor factors and the heavy quark propagator.
At least, there exist bound state corrections in Eq.(14) which are related to Eqs.(3) and (4).

Notice that there are several energy scales which can characterize the heavy
quarkonium: the hard momentum scale $m_Q$ (the mass of heavy quark),
the soft momentum scale $m_Qv_Q$ ($v_Q$ is the heavy quark velocity in
the bound state) and the ultrasoft momentum scale
$m_Qv_Q^2$. We assume that the heavy quarkonium is a nonrelativistic
system. This means that the following inequalities occur:
$m_Q\gg m_Qv_Q\gg m_Qv_Q^2$, $m_Q\gg \Lambda_{QCD}$, which we can
exploit in the further study of the production amplitude (14).
The expansion of basic factors over relative momenta ${\bf p}$
and ${\bf q}$ up to terms of the second order has the following
form:
\begin{equation}
\frac{1}{k^2}=\frac{1}{Z_1}\left[1-\frac{p^2+q^2}{Z_1}+\frac{2[\rho_2(rp)+
\eta_2(rq)]}{Z_1}+\frac{4[\rho_2^2(rp)^2+\eta_2^2(rq)^2]}{Z_1^2}\right],
\end{equation}
\begin{displaymath}
\frac{1}{(r-q_1)^2-m_1^2}=\frac{1}{Z_2}\left[1-\frac{q^2}{Z_2}+\frac{2(rq)}
{Z_2}+\frac{4(rq)^2}{Z_2^2}\right],~~Z_1=\eta_2\rho_2r^2,
\end{displaymath}
\begin{displaymath}
Z_2=r^2-2\rho_1(rQ)+\rho^2_1M_{\cal P}^2-m_1^2,~~r=P+Q,
\end{displaymath}
where we have chosen the center-of-mass frame with the condition
${\bf P}+ {\bf Q}=0$; $r_0=\sqrt{s}$ is the total energy in the
electron-positron annihilation. In the leading order, when one
neglects relativistic and bound state corrections, one obtains
$Z_1=(1-\kappa)^2s$, $Z_2=(1-\kappa)s$, $\kappa$ = $ m_1/M_0$ = $m_1/(m_1+m_2)$.
So, the expansions (15) are well defined. Substituting (15) into
Eq.(14) we have to combine the factors of the second degree over
$p$ and $q$. After averaging the obtained expression over the angle
variables with the account of orthogonality conditions (3) and
(4) \cite{apm2005}, we can present the necessary correction in the
integral form $\int d{\bf p}{\bf p}^2 \bar\Psi_0^{\cal V}({\bf
p})$. Moreover, we performed the expansion of the amplitude
(14) at ${\bf p}={\bf q}=0$ over the binding energies $W_{\cal P}$
and $W_{\cal V}$ in the linear approximation in such a way that
relativistic and bound state contributions can exist separately
as follows:
\begin{equation}
{\cal M}_1(p_-,p_+,P,Q)=\frac{8\pi^2\alpha\alpha_s}{3}\bar v(p_+)\gamma^\beta u(p_-)
\int\frac{d{\bf p}}{(2\pi)^3}\bar\Psi_0^{\cal V}({\bf p})
\int\frac{d{\bf q}}{(2\pi)^3}\bar\Psi_0^{\cal P}({\bf q})\times
\end{equation}
\begin{displaymath}
\times \epsilon_{\sigma\rho\lambda\beta}v_1^\sigma
v_2^\rho\tilde\epsilon^{\ast~\lambda}
\frac{M_0}{s^3(1-\kappa)^3}\Biggl\{1+\frac{2W_{\cal P}}{M_0(1-\kappa)}
(1-2\kappa+\kappa\frac{M_0^2}{s})+\frac{W_{\cal V}}{M_0(1-\kappa)}
(2-3\kappa-2\kappa\frac{M_0^2}{s})+
\end{displaymath}
\begin{displaymath}
\frac{{\bf p}^2}{72M_0^2\kappa^2(1-\kappa)^2}\left[-9+8\kappa\left(4+\frac{M_0}{\sqrt{s}})-
8\kappa^2(1-14\frac{M_0^2}{s}+10\frac{M_0^4}{s^2}\right)\right]+
\frac{{\bf q}^2}{72M_0^2\kappa^2(1-\kappa)^2}\times
\end{displaymath}
\begin{displaymath}
\left[-9+16\kappa(1+\frac{M_0}{\sqrt{s}}-
16\kappa^3\frac{M_0^2}{s}\left(3-\frac{4M_0}{\sqrt{s}}+\frac{14M_0^2}{s}\right)+
32\kappa^2\frac{M_0}{\sqrt{s}}\left(-1+\frac{M_0}{\sqrt{s}}-\frac{2M_0^2}{s}+
\frac{2M_0^3}{s^{3/2}}\right)\right]\Biggr\}.
\end{displaymath}

\begin{figure}[htbp]
\vspace{-4mm}
\begin{center}
$e^++e^- \to J/\Psi+\eta_c$
\end{center}
\centering
\includegraphics{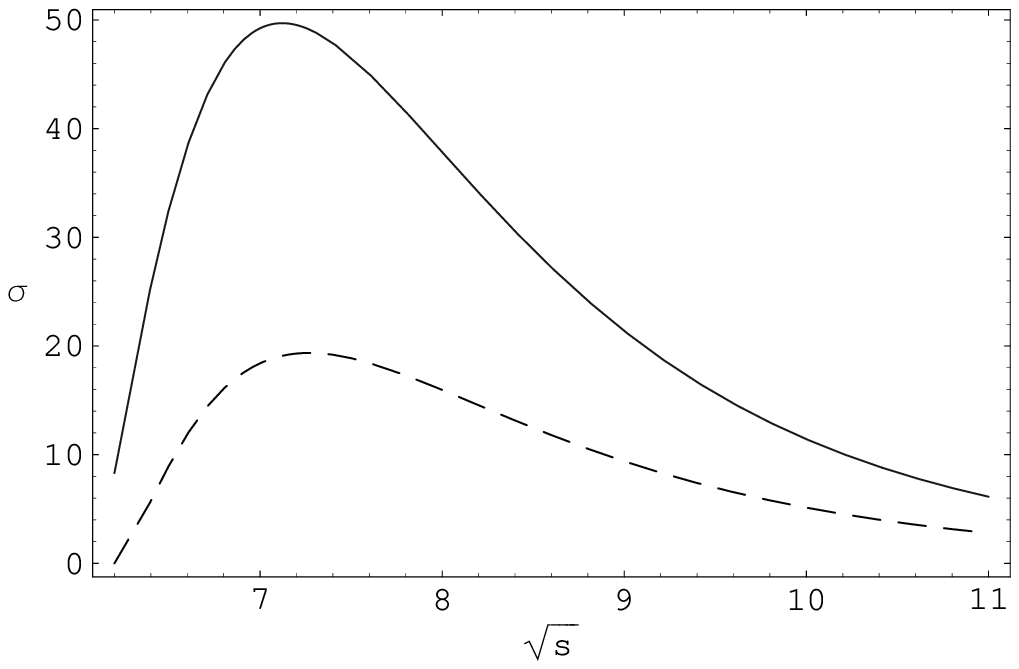}
\vspace{-3mm}
\begin{center}
$e^++e^- \to B^+_c+B^{\ast -}_c$
\end{center}
\centering
\includegraphics{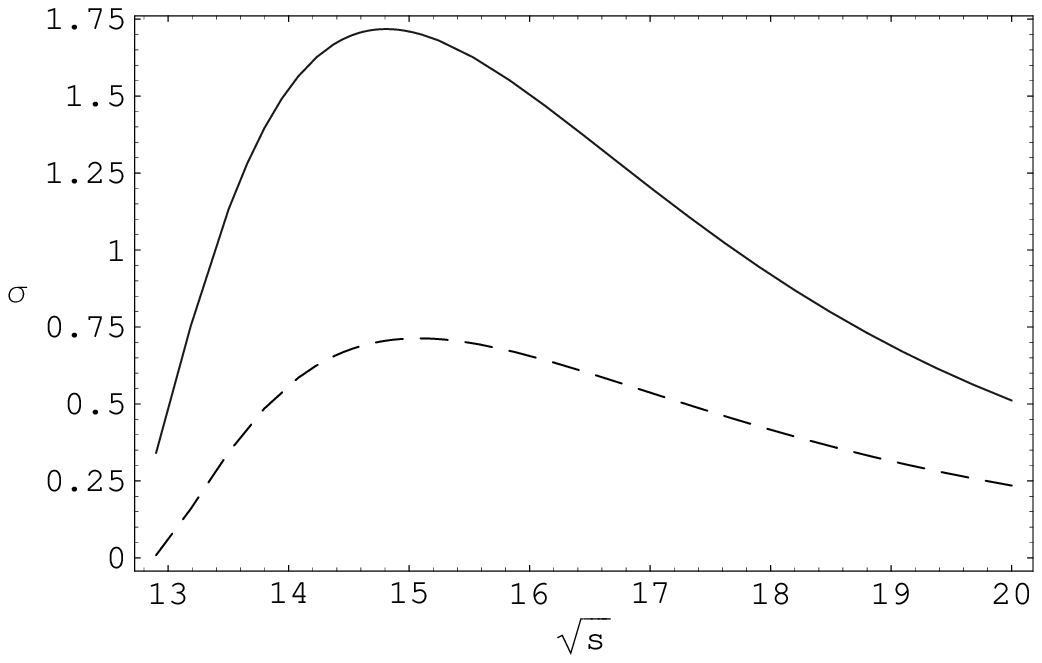}
\vspace{-3mm}
\begin{center}
$e^++e^- \to \Upsilon+\eta_b$
\end{center}
\centering
\includegraphics{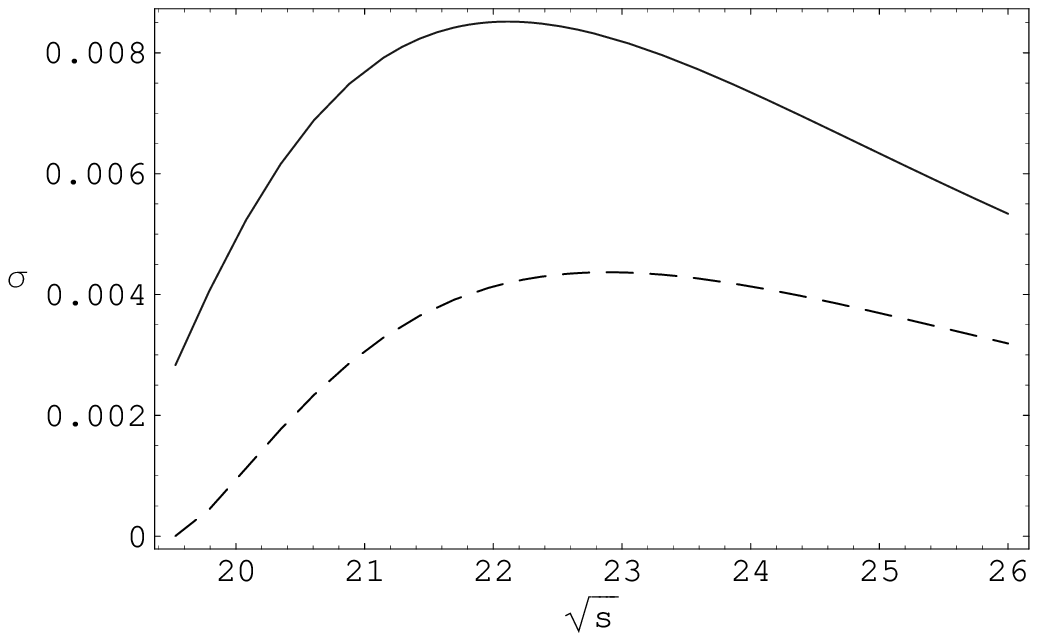}
\vspace{-2mm}
\caption{The cross section in fb of $e^+e^-$
annihilation into a pair of doubly heavy mesons with opposite
charge parity as a function of the center-of-mass energy $\sqrt{s}$
(solid line). The dashed line shows the nonrelativistic result
without bound state and relativistic corrections.}
\end{figure}

Similar expressions occur for the other amplitudes contributing to the
production process:

\begin{equation}
{\cal M}_2(p_-,p_+,P,Q)=\frac{8\pi^2\alpha\alpha_s}{3}\bar v(p_+)\gamma^\beta u(p_-)
\int\frac{d{\bf p}}{(2\pi)^3}\bar\Psi_0^{\cal V}({\bf p})
\int\frac{d{\bf q}}{(2\pi)^3}\bar\Psi_0^{\cal P}({\bf q})\times
\end{equation}
\begin{displaymath}
\times\frac{M_0}{s^3(1-\kappa)^3}\Biggl\{1+\frac{2W_{\cal P}}{M_0}
(1-\frac{M_0^2}{s})-\frac{W_{\cal V}}{M_0\kappa(1-\kappa)}
(1+\kappa(-2-2\kappa+\kappa^2+2(1-\kappa)\frac{M_0^2}{s})+
\end{displaymath}
\begin{displaymath}
\frac{{\bf p}^2}{72M_0^2\kappa^2(1-\kappa)^2}\Biggl[-9+40\kappa\left(1+
\frac{M_0}{\sqrt{s}})+
4\kappa^2(1-18\frac{M_0}{\sqrt{s}}+24\frac{M_0^2}{s}-16\frac{M_0^3}{s^{3/2}}+
16\frac{M_0^4}{s^2}\right)-
\end{displaymath}
\begin{displaymath}
-4\kappa^3\left(7-4\frac{M_0}{\sqrt{s}}+28\frac{M_0^2}{s}-16\frac{M_0^3}{s^{3/2}}+
56\frac{M_0^4}{s^2}\right)\Biggr]+
\frac{{\bf q}^2}{72M_0^2\kappa^2(1-\kappa)^2}\times
\end{displaymath}
\begin{displaymath}
\times\left[-9-8\kappa(-8+\frac{M_0}{\sqrt{s}})-8\kappa^2\left(5-2\frac{M_0}{\sqrt{s}}-
2\frac{M_0^2}{s}+4\frac{M_0^4}{s^2}\right)\right]\Biggr\}\epsilon_{\sigma\rho\lambda
\beta}v_1^\sigma v_2^\rho\tilde\epsilon^{\ast~\lambda},
\end{displaymath}
\begin{equation}
{\cal M}_3(p_-,p_+,P,Q)=\frac{8\pi^2\alpha\alpha_s}{3}\bar v(p_+)\gamma^\beta u(p_-)
\int\frac{d{\bf p}}{(2\pi)^3}\bar\Psi_0^{\cal V}({\bf p})
\int\frac{d{\bf q}}{(2\pi)^3}\bar\Psi_0^{\cal P}({\bf q})\times
\end{equation}
\begin{displaymath}
\times\frac{M_0}{s^3\kappa^3}\Biggl\{1+\frac{W_{\cal P}}{M_0\kappa}
\left(-1-\frac{2M_0^2}{s}+2\kappa(2+\frac{M_0^2}{s})\right)+\frac{W_{\cal V}}
{M_0\kappa}\left(-2+\frac{2M_0^2}{s}(1-\kappa)+2\kappa+\kappa^2\right)+
\end{displaymath}
\begin{displaymath}
\frac{{\bf p}^2}{72M_0^2\kappa^2(1-\kappa)^2}\Biggl[7-16\frac{M_0}{\sqrt{s}}-
16\frac{M_0^2}{s}-160\frac{M_0^4}{s^2}+4\kappa\left(5+10\frac{M_0}{\sqrt{s}}+
36\frac{M_0^2}{s}-16\frac{M_0^3}{s^{3/2}}+136\frac{M_0^4}{s^2}\right)+
\end{displaymath}
\begin{displaymath}
+8\kappa^2\left(4-3\frac{M_0}{\sqrt{s}}+
30\frac{M_0^2}{s}-16\frac{M_0^3}{s^{3/2}}+76\frac{M_0^4}{s^2}\right)
+4\kappa^3\left(-1-12\frac{M_0}{\sqrt{s}}+28\frac{M_0^2}{s}-16\frac{M_0^3}
{s^{3/2}}+56\frac{M_0^4}{s^2}\right)\Biggr]+
\end{displaymath}
\begin{displaymath}
+\frac{{\bf q}^2}{72M_0^2\kappa^2(1-\kappa)^2}\Biggl[15+8\frac{M_0}{\sqrt{s}}
+16\frac{M_0^2}{s}-32\frac{M_0^4}{s^2}
+8\kappa\left(2-3\frac{M_0}{\sqrt{s}}-4\frac{M_0^2}{s}+8\frac{M_0^4}{s^2}\right)-
\end{displaymath}
\begin{displaymath}
-8\kappa^2\left(5-2\frac{M_0}{\sqrt{s}}-2\frac{M_0^2}{s}+
4\frac{M_0^4}{s^2}\right)\Biggr]\Biggr\}\epsilon_{\sigma\rho\lambda\beta}v_1^\sigma
v_2^\rho\tilde\epsilon^{\ast~\lambda},
\end{displaymath}
\begin{equation}
{\cal M}_4(p_-,p_+,P,Q)=\frac{8\pi^2\alpha\alpha_s}{3}\bar v(p_+)\gamma^\beta u(p_-)
\int\frac{d{\bf p}}{(2\pi)^3}\bar\Psi_0^{\cal V}({\bf p})
\int\frac{d{\bf q}}{(2\pi)^3}\bar\Psi_0^{\cal P}({\bf q})\times
\end{equation}
\begin{displaymath}
\times\frac{M_0}{s^3\kappa^3}\Biggl\{1+\frac{W_{\cal P}}{M_0\kappa}
\left(-2+\frac{2M_0^2}{s}+2\kappa(2-\frac{M_0^2}{s})\right)+\frac{W_{\cal V}}
{M_0\kappa}\left(-1-\frac{2M_0^2}{s}+\kappa(3+\frac{2M_0^2}{s})\right)+
\end{displaymath}
\begin{displaymath}
\frac{{\bf p}^2}{72M_0^2\kappa^2(1-\kappa)^2}\Biggl[15+8\frac{M_0}{\sqrt{s}}+
16\frac{M_0^2}{s}-32\frac{M_0^4}{s^2}+8\kappa\left(-2-\frac{M_0}{\sqrt{s}}-
4\frac{M_0^2}{s}+8\frac{M_0^4}{s^2}\right)-
\end{displaymath}
\begin{displaymath}
-8\kappa^2\left(1-2\frac{M_0^2}{s}+4\frac{M_0^4}{s^2}\right)\Biggr]+
\frac{{\bf q}^2}{72M^2\kappa^2(1-\kappa)^2}\Biggl[7-16\frac{M_0}{\sqrt{s}}
-16\frac{M_0^2}{s}-160\frac{M_0^4}{s^2}+
\end{displaymath}
\begin{displaymath}
+16\kappa\left(-1+3\frac{M_0}{\sqrt{s}}+
5\frac{M_0^2}{s}-4\frac{M_0^3}{s^{3/2}}+34\frac{M_0^4}{s^2}\right)-
16\kappa^2\frac{M_0}{\sqrt{s}}\left(2+7\frac{M_0}{\sqrt{s}}-8\frac{M_0^2}{s}+
38\frac{M_0^3}{s^{3/2}}\right)+
\end{displaymath}
\begin{displaymath}
+16\kappa^3\frac{M_0^2}{s}\left(3-4\frac{M_0}{\sqrt{s}}+14\frac{M_0^2}{s}\right)
\Biggr]\Biggr\}\epsilon_{\sigma\rho\lambda\beta}v_1^\sigma
v_2^\rho\tilde\epsilon^{\ast~\lambda}.
\end{displaymath}

It is important to emphasize that we carried out formal expansions of the 
integrands in the amplitudes ${\cal M}_i$ (16)-(19) over quantities
${\bf p^2}/m_{1,2}^2$, ${\bf q^2}/m_{1,2}^2$ assuming that the relative
momenta of heavy quarks in the bound state are small as compared with
their masses. This means that the formally divergent integrals 
$\int d{\bf p}{\bf p^2}\bar\Psi({\bf p})$
and $\int d{\bf q}{\bf q^2}\bar\Psi({\bf q})$ have to be defined by a suitable
regularization procedure. Their numerical values will be directly determined by
the properties of the wave functions of the heavy quark bound states.

\section{Cross section of pseudoscalar and vector doubly heavy mesons
production}

Using the relativistic amplitudes ${\cal M}_i$ obtained above, we can calculate the
production cross section for doubly heavy mesons in $e^+e^-$ annihilation
as a function of the center-of-mass energy $\sqrt{s}$. For this purpose, the sum
of amplitudes ${\cal M}_i$ (16)-(19) can be represented as follows:
\begin{equation}
{\cal M}(e^+e^-\to {\cal P}+{\cal
V})=\frac{16}{27}\pi^2\alpha\frac{M_0\sqrt{4M_{\cal P} M_{\cal
V}}}{s^6\kappa^5(1-\kappa)^5}\bar\Psi_0^{\cal
V}(0)\bar\Psi_0^{\cal P}(0)\times
\end{equation}
\begin{displaymath}
\times\bar v(p_+)\gamma^\beta u(p_-)\epsilon_{\sigma\rho\lambda\beta}v_1^\sigma
v_2^\rho\tilde\epsilon^{\ast~\lambda}\left[\kappa^3Q_1\alpha_{s2}T_1+
(1-\kappa)^3Q_2\alpha_{s1}T_2\right],
\end{displaymath}
where $\alpha_{s1}=\alpha_s(4m_1^2)$,
$\alpha_{s2}=\alpha_s(4m_2^2)$, $Q_{1,2}$ are the electric
charges of heavy quarks, $\Psi_0^{\cal V,P}(0)$ are the wave
functions for the relative motion of heavy quarks in the vector
and pseudoscalar meson at the origin in its rest frame,
\begin{equation}
T_1=72\kappa^2(1-\kappa)^2+\frac{\left\langle{\bf p}^2\right\rangle}{M_0^2}
\Bigl[-9+36\kappa-2\kappa^2-
14\kappa^3+\frac{M_0}{\sqrt{s}}4\kappa(6-9\kappa+2\kappa^2)+
\end{equation}
\begin{displaymath}
+\frac{M_0^2}{s}56\kappa^2(1-\kappa)+\frac{M_0^3}{s^{3/2}}32\kappa^2(\kappa-1)+
\frac{M_0^4}{s^2}16\kappa^2(1-7\kappa)\Bigr]+
\frac{\left\langle{\bf q}^2\right\rangle}{M_0^2}\Bigl[-9+40
\kappa-20\kappa^2+
\end{displaymath}
\begin{displaymath}
+\frac{M_0}{\sqrt{s}}4\kappa(1-2\kappa)
+\frac{M_0^2}{s}24\kappa^2(1-\kappa)+\frac{M_0^3}{s^{3/2}}32\kappa^2(\kappa-1)+
\frac{M_0^4}{s^2}16\kappa^2(1-7\kappa)\Bigr]+
\end{displaymath}
\begin{displaymath}
+\frac{W_{\cal V}}{M_0}36\kappa\left[1-5\kappa^2+6\kappa^3-\kappa^4+\frac{M_0^2}
{s}2\kappa(1-\kappa)^2\right]
+\frac{W_{\cal P}}{M_0}72\kappa^2\left[2-5\kappa+3\kappa^2-\frac{M_0^2}
{s}(1-\kappa)^2\right],
\end{displaymath}
\begin{equation}
T_2=72\kappa^2(1-\kappa)^2+\frac{\left\langle{\bf p}^2\right\rangle}{M_0^2}\Bigl[11+2\kappa-20\kappa^2-
2\kappa^3+\frac{M_0}{\sqrt{s}}4(-1+4\kappa+3\kappa^2-6\kappa^3)+
\end{equation}
\begin{displaymath}
+\frac{M_0^2}{s}56\kappa(1-\kappa)^2-\frac{M_0^3}{s^{3/2}}32\kappa(\kappa-1)^2+
\frac{M_0^4}{s^2}16(-6+19\kappa-20\kappa^2+7\kappa^3)\Bigr]+
\frac{\left\langle{\bf q}^2\right\rangle}{M_0^2}\Bigl[11-20\kappa^2+
\end{displaymath}
\begin{displaymath}
+\frac{M_0}{\sqrt{s}}4(-1+3\kappa-2\kappa^2)
+\frac{M_0^2}{s}24\kappa(1-\kappa)^2-\frac{M_0^3}{s^{3/2}}32\kappa(\kappa-1)^2+
\frac{M_0^4}{s^2}16(-6+19\kappa-20\kappa^2+7\kappa^3)\Bigr]+
\end{displaymath}
\begin{displaymath}
+\frac{W_{\cal V}}{M_0}36\kappa(-3+11\kappa-12\kappa^2+3\kappa^3+\kappa^4)
+\frac{W_{\cal P}}{M_0}36\kappa(-3+14\kappa-19\kappa^2+8\kappa^3),
\end{displaymath}
where ${\left\langle{\bf p}^2\right\rangle}$,
${\left\langle{\bf q}^2\right\rangle}$ are quantities determining
the numerical values of relativistic effects connected with the internal motion
of the heavy quarks in vector and pseudoscalar doubly heavy mesons. Note that they are
not equal to the matrix elements $\int d{\bf p}\bar\Psi_0^{\cal V}({\bf p})
{\bf p}^2\Psi_0^{\cal V}({\bf p})$
and $\int d{\bf q}\bar\Psi_0^{\cal P}({\bf q}){\bf q}^2\Psi_0^{\cal P}
({\bf q})$ and are discussed at the end of Sec. III.

\begin{figure}[htbp]
\vspace{-4mm}
\begin{center}
$e^++e^- \to J/\Psi+\eta'_c$
\end{center}
\centering
\includegraphics{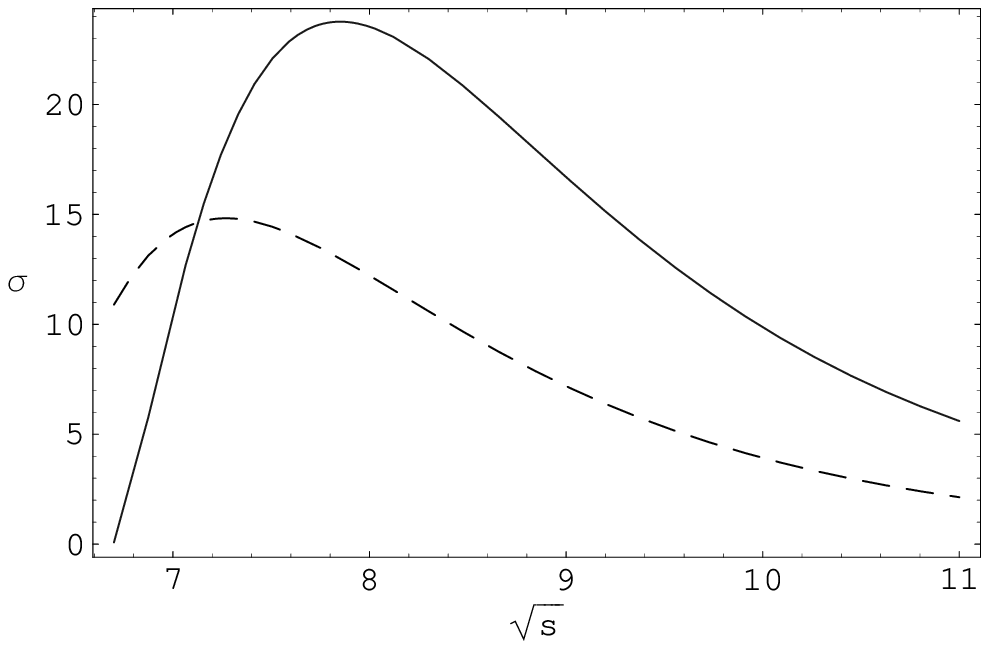}
\vspace{-3mm}
\begin{center}
$e^++e^- \to \Psi'+\eta_c$
\end{center}
\centering
\includegraphics{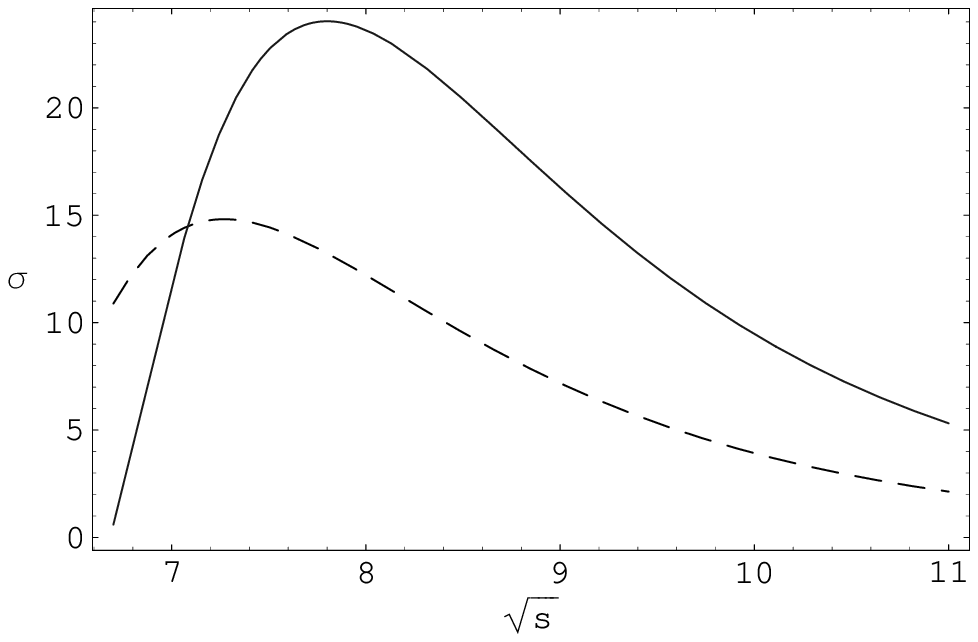}
\vspace{-3mm}
\begin{center}
$e^++e^- \to \Psi'+\eta'_c$
\end{center}
\centering
\includegraphics{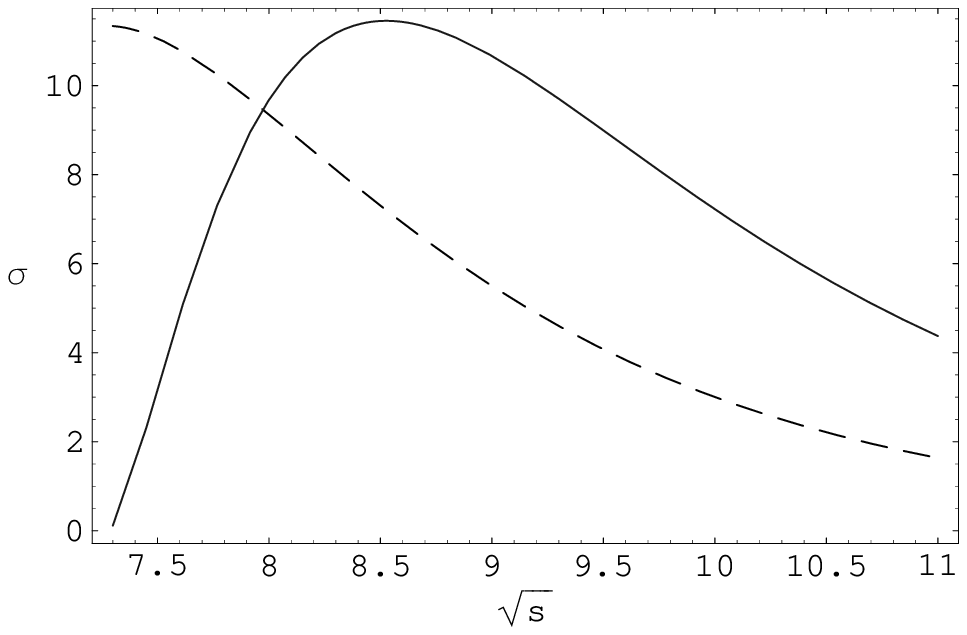}
\vspace{-2mm}
\caption{The cross section in fb of $e^+e^-$
annihilation into a pair of $S$-wave doubly charm heavy mesons with opposite
charge parity as a function of the center-of-mass energy $s$
(solid line). The dashed line shows the nonrelativistic result
without bound state and relativistic corrections.}
\end{figure}

Performing standard algebraic calculations with the squared modulus
of the amplitude $|{\cal M}|^2$ by the use of the system Form \cite{Form}, we
obtain the following differential cross section:
\begin{equation}
\frac{d\sigma}{d\cos\theta}=\frac{4\pi^3\alpha^2M_0^2|\Psi_0^{\cal
V}(0)|^2 |\Psi_0^{\cal P}(0)|^2}{729M_{{\cal V}}M_{{\cal
P}}s^8\kappa^{10}(1-\kappa)^{10}}
\left[\kappa^3Q_1\alpha_{s2}T_1+(1-\kappa)^3Q_2\alpha_{s1}T_2\right]^2\times
\end{equation}
\begin{displaymath}
\times\left\{\frac{[s^2-(M_{{\cal V}}+M_{{\cal
P}})^2][s^2-(M_{{\cal V}}-M_{{\cal P}})^2]}
{s^4}\right\}^{3/2}\left(1+\cos^2\theta\right),
\end{displaymath}
where $\theta$ is the angle between the momenta of the incoming
lepton and the outgoing doubly heavy meson. The total cross
section for the exclusive production of pseudoscalar and vector
doubly heavy mesons in $e^+e^-$ annihilation is then given by the
following expression:
\begin{equation}
\sigma(s)=\frac{32\pi^3\alpha^2M_0^2|\Psi_0^{\cal V}(0)|^2
|\Psi_0^{\cal P}(0)|^2}{2187M_{{\cal V}}M_{{\cal
P}}s^8\kappa^{10}(1-\kappa)^{10}}
\left[\kappa^3Q_1\alpha_{s2}T_1+(1-\kappa)^3Q_2\alpha_{s1}T_2\right]^2\times
\end{equation}
\begin{displaymath}
\times\left\{\left[1-\frac{(M_{{\cal V}}+M_{{\cal P}})^2}{s^2}\right]
\left[1-\frac{(M_{{\cal V}}-M_{{\cal P}})^2}{s^2}\right]\right\}^{3/2},
\end{displaymath}
If we neglect relativistic and bound state effects in Eq.(24), that is
$\left\langle{\bf p}^2\right\rangle$ = $\left\langle{\bf q}^2\right\rangle$
=0, $M_{{\cal P}}=M_{{\cal V}}=M_0$, then we obtain $\sigma_{NR}(s)$ coinciding
with the analytical result of Ref.\cite{VK}. To estimate how the relativistic
and bound state corrections can change the nonrelativistic cross section $\sigma_{NR}(s)$,
we have used definite numerical values of a number of parameters entering
in Eqs. (21)-(24). They can be determined on the basis of the relativistic
quark model \cite{rqm1,rqm2,rqm3} as we shall discuss in the following.

In the quasipotential approach the bound
states of heavy quarks are described by the Schr\"odinger type equation
\cite{rqm1}
\begin{equation}
\label{quas}
{\left(\frac{b^2(M)}{2\mu_{R}}-\frac{{\bf p}^2}{2\mu_{R}}\right)
\Psi_0({\bf p})} =\int\frac{d{\bf q}}{(2\pi)^3}V({\bf p,q},M)
\Psi_0({\bf q}),
\end{equation}
where the relativistic reduced mass is
\begin{equation}
\mu_{R}=\frac{E_1E_2}{E_1+E_2}=\frac{M^4-(m^2_1-m^2_2)^2}{4M^3},
\end{equation}
and the particle energies $E_1$, $E_2$ are given by
\begin{equation}
E_1=\frac{M^2-m_2^2+m_1^2}{2M}, \quad E_2=\frac{M^2-m_1^2+m_2^2}{2M}.
\end{equation}
Here $M=E_1+E_2$ is the bound state mass,
$m_{1,2}$ are the masses of heavy quarks ($Q_1$ and $Q_2$) which form
the meson, and ${\bf p}$  is their relative momentum.
In the center of mass system the relative momentum squared on the
energy surface $M=E_1+E_2$ reads
\begin{equation}
{b^2(M)}=\frac{[M^2-(m_1+m_2)^2][M^2-(m_1-m_2)^2]}{4M^2}.
\end{equation}
For small binding energies it is approximately equal to $2\mu_R W$.
Notice that the kernel $V({\bf p,q},M)$ in Eq.(\ref{quas}) is the quasipotential
operator of the quark-antiquark interaction.
The construction of the quark interaction operator is discussed permanently
during the last decades \cite{UFN1,QWG,Bali,LSG}. A new stage in solving this problem
came when nonrelativistic field theories were introduced for the study
of nonrelativistic bound states \cite{CL,Patrick,RMP}.
Within an effective field theory
(NRQCD) the quark-antiquark potential was constructed in Refs.\cite{RMP,NR1,NR2}
by the perturbation theory improved renormalization group resummation
of large logarithms. 
On the other hand, in the quasipotential quark model the kernel $V({\bf
p},{\bf q},M)$ is constructed phenomenologically with the help of the
off-mass-shell scattering amplitude, projected onto the positive
energy states. The heavy quark-antiquark potential with the account of
retardation effects and the one-loop radiative corrections can then be presented
in the form of a sum of spin-independent and spin-dependent parts.
The explicit expression for it is given in Refs.\cite{rqm2,rqm3}.
Taking into account the accuracy of the calculation of relativistic
corrections to the cross section (24), we can use for the description
of the bound system $(Q_1 \bar Q_2)$ the following simplified interaction
operator in the coordinate representation:
\begin{equation}
\tilde V(r)=-\frac{4}{3}\frac{\alpha_s(\mu^2)}{r}+Ar+B,
\end{equation}
where the parameters of the linear potential $A=0.18~GeV^2$, $B=-0.3~GeV$,
\begin{equation}
\alpha_s(\mu^2)=\frac{4\pi}{\beta_0\ln(\mu^2/\Lambda^2)},~~~
\beta_0=11-\frac{2}{3}n_f.
\end{equation}
Here $n_f=3$ is the number of flavours and
$\mu=\frac{2m_1m_2}{(m_1+m_2)}$ is a renormalization scale and 
$\Lambda=0.168$ GeV. All the parameters of the model like quark
masses, parameters of the linear confining potential $A$ and $B$,
mixing coefficient $\varepsilon$ and anomalous chromomagnetic
quark moment $\kappa$ entering in the quasipotential $V({\bf
p},{\bf q},M)$ were fixed from the analysis of heavy quarkonium
masses \cite{rqm1,rqm2,rqm3,rqm4} and radiative decays
\cite{rqm2}. Solving the Schr\"odinger-like quasipotential
equation with the operator (29) we obtain an initial
approximation for the bound state wave functions in the case of
$(c\bar c)$, $(\bar b c)$ and $(\bar b b)$ systems. For numerical
estimations of relativistic and bound state effects in the
production of heavy mesons in the $e^+e^-$ annihilation we need
the values of the wave functions at the origin, the bound state
energy and the parameter of relativistic effects: $\int{\bf
p}^2\bar{\Psi}_0^{\cal P,V}({\bf p})d{\bf p}/(2\pi)^3$. This
integral diverges at high momentum. Using the Schr\"odinger-like
equation (25), we can express it through the value of the
potential $V(r)$ at the origin. Different regularizations for it
are discussed and already applied to a number of tasks concerning the
production and decay of bound states
\cite{Labelle,aKM,KM,CMY}. A recent discussion of
the method for the calculation of an order-$v^2$ matrix element
is presented in Ref.\cite{BKL}. To estimate the numerical value of
relativistic corrections in the production processes, we follow the
prescription of dimensional regularization, where the scaleless
momentum integral $\int V({\bf p}-{\bf q})\Psi({\bf
q})\frac{d^d{\bf q}}{(2\pi)^d}\frac{d^d{\bf p}}{(2\pi)^d}$
related to the problem vanishes \cite{aKM,Labelle,CMY}. Then we
can express the necessary quantity in the form:
\begin{equation}
\left\langle {\bf p}^2\right\rangle_{DR}\equiv \frac{1}{\Psi(0)}\int \frac{d^d{\bf
p}}{(2\pi)^d}{\bf p}^2\bar\Psi_0({\bf p})=2\mu_R \tilde W+2\mu_R |B|.
\end{equation}
Let us note, that in the case of the Coulomb interaction
in QED the dependence of the relativistic parameter (31) on the
principal quantum number $n$ is known in the analytical form
\cite{MF1999}. For the quark bound states
the dependence on the quantum number $n$ is determined numerically.
Solving the quasipotential equation (25) with the
potential (29) \cite{FFS}, we obtain the energy spectrum $\tilde W$ of the
heavy quark system and the numerical values of the
parameter (31) for the bound states $(\bar cc)$, $(\bar bb)$ and
$(\bar b c)$ which are presented in Table I. Heavy quark symmetry
predicts that the wave functions of the vector and pseudoscalar
states are different due to corrections of order $v_Q^2$. The
analogous statement is valid for the parameter $\left\langle {\bf
p}^2\right\rangle$. Neglecting in this study the corrections of
order $O(v^4_Q)$ in the production rates we write in Table I equal values 
of $\left\langle {\bf p}^2\right\rangle$ for $V$- and $P$-mesons.
The theoretical uncertainty of the obtained value of the parameter $\left\langle {\bf p}^2\right\rangle$ in Table I is determined by
perturbative and nonperturbative corrections to the quasipotential
\cite{rqm1,rqm2} and is not exceeding $30\%$. We presented in Table I
the estimation of the pseudoscalar and vector wave functions
at the origin including effects of order $O(v_Q^2)$. For this aim
first of all, we obtained the central equal values of $\Psi_{\cal V}(0)$ and $\Psi_{\cal P}(0)$ solving the Schr\"odinger-like equation (25). Then
we took into account the spin-dependent corrections of order $O(v_Q^2)$
to the wave functions at the origin which were calculated in the framework of nonrelativistic QCD (see Refs.\cite{NR1,W1,W2,W3}).

\begin{table}
\caption{\label{t1} Basic parameters of the relativistic quark model}
\bigskip
\begin{ruledtabular}
\begin{tabular}{|c|c|c|c|c|c|c|}   \hline
State  & Particle & Mass,~$GeV$  & $\alpha_s$ & Bound state &
$\Psi(0),~GeV^{3/2}$ &$\left\langle{\bf p}^2\right\rangle_{DR}$,   \\
$n^{2S+1}L_J$ &  &\cite{rqm2,PDG}  &  & energy,~$GeV$ &  & $GeV^2$  \\ \hline $1^1S_0$ & $\eta_c$ & 2.980 & 0.314 &-0.120
&0.28   & 0.6  \\ \hline $1^3S_1$ & $J/\Psi$ & 3.097 &
0.314 &-0.003  &0.26   &0.6  \\  \hline $2^1S_0$ &
$\eta'_c$ &3.594  &0.314  & 0.494 & 0.25  & 1.4   \\
\hline $2^3S_1$ & $\Psi'$ &3.686  &0.314  & 0.586 &0.23
&1.4  \\ \hline $1^1S_0$ & $B_c^+$ & 6.270 & 0.265 &-0.160
&0.36 &0.7  \\  \hline $1^3S_1$ & $B^{\ast -}_c$ & 6.332 &
0.265 &-0.098 &0.34   &0.7  \\  \hline $2^1S_0$ &
$B'^+_c$ & 6.835 &0.265  & 0.405 & 0.34 &1.8  \\  \hline
$2^3S_1$ & $B'^{\ast -}_c$ &6.881  &0.265  &0.451  &0.32  &1.8
  \\  \hline $1^1S_0$ & $\eta_b$ & 9.400 & 0.207 & -0.360  &
0.55   & 0.9  \\ \hline $1^3S_1$ & $\Upsilon$ & 9.460 &
0.207 &-0.300 & 0.53 & 0.9   \\  \hline $2^1S_0$ &
$\eta'_b$ & 9.993 &0.207  &0.233 & 0.45 &2.9  \\
\hline $2^3S_1$ & $\Upsilon'$ &10.023  &0.207  &0.263 & 0.43
&2.9   \\  \hline
\end{tabular}
\end{ruledtabular}
\end{table}

\section{Summary and discussion}

We have investigated the relativistic and bound state effects in the production of doubly heavy
mesons on the basis of the perturbative QCD and the
relativistic quasipotential
quark model. Using the factorization hypothesis we keep systematically
all relativistic corrections of the second order in the relative
velocity of heavy quarks and bound state effects in the production of the $S$-wave pseudoscalar and vector mesons from $e^+e^-$ annihilation.

Let us summarize the basic peculiarities of the calculation performed
above.

1. We obtain the cross sections for the production of a pair of $S$-wave doubly heavy mesons with opposite charge parity
containing $b$ and $c$ quarks from $e^+e^-$ annihilation.

2. All possible sources of relativistic corrections including the
transformation factors for the two quark bound state wave function
have been taken into account.

3. We have investigated the role of relativistic and bound state effects in the
total production cross sections using predictions of the relativistic
quark model for a number of parameters entering in the obtained
analytical expressions.

The results of our calculation of the cross section (24) presented in Figs.2-3 evidently show that only the relativistic analysis of the
production processes can give reliable theoretical predictions for
the comparison with the experimental data.
It follows from Fig.3 that with the growth of the quantum number
$n$ the nonrelativistic approximation doesn't work near the production threshold
because the omitted
terms in this case have the same order of the magnitude as the basic terms.

\begin{table}
\caption{\label{t1} Comparison of theoretical predictions with experimental
data.}
\bigskip
\begin{ruledtabular}
\begin{tabular}{|c|c|c|c|c|c|c|c|}   \hline
State  & $\sigma_{BaBar}\times$ & $\sigma_{Belle}\times $ &
$\sigma$ $(fb)$ &$\sigma_{NRQCD}$& $\sigma$ $(fb)$ &$\sigma$ $(fb)$ &Our result   \\
$H_1H_2$   &$ Br_{H_2\to charged\ge 2}$ & $
   Br_{H_2\to charged\ge 2}$ & \cite{BLL} &$(fb)$ \cite{BL1}& \cite{Chao1}  &\cite{BL1}
   & $(fb)$  \\  
   &  $(fb)$ \cite{BaBar} & $(fb)$ \cite{Belle} &   &  &  &    &    \\   \hline
$\Psi(1S)\eta_c(1S)$ & $17.6\pm 2.8^{+1.5}_{-2.1}$ & $25.6\pm 2.8\pm 3.4$ &
26.7 &  3.78& 5.5 & 7.4  & 7.8  \\  \hline
$\Psi(2S)\eta_c(1S)$ &  & $16.3\pm 4.6\pm 3.9$ &
16.3 &  1.57& 3.7 & 6.1  & 6.7 \\  \hline
$\Psi(1S)\eta_c(2S)$ & $16.4\pm 3.7^{+2.4}_{-3.0}$ & $16.5\pm 3.\pm 2.4$ &
26.6 &  1.57 &3.7 &7.6 & 7.0  \\  \hline
$\Psi(2S)\eta_c(2S)$ &  & $16.0\pm 5.1\pm 3.8$ &
14.5 &  0.65& 2.5 &5.3 &5.4   \\  \hline
\end{tabular}
\end{ruledtabular}
\end{table}

As we have already mentioned, the experimental results for the production
of $J/\Psi+\eta_c$ mesons measured at the Belle and BaBar experiments
differ from theoretical calculations in the framework of NRQCD.
The experimental data on the production cross sections
of a pair of $S$-wave charm mesons are presented in Table II.
The numerical value for the cross section of
$J/\Psi+\eta_c$ production at $\sqrt{s}=10.6$ GeV, obtained on the
basis of Eq.(24) amounts to the value 7.8 fb without the inclusion
of QED effects. In this case relativistic and bound
state corrections increase our nonrelativistic result by a factor
$2.2$ (cf. dashed lines in Figs.2,3).
Accounting slightly different values of several parameters
used in our model in the comparison with Ref.\cite{BL1} (the mass of
$c$ quark, the binding energies $W_{\cal P,V}$), we find
agreement of our calculations with the results of Ref.\cite{BL1}
for the production of the charmonium states, if relativistic corrections
are taken into account (see the seventh column of Table II). Taking
in mind also the calculation \cite{ZGC}, which includes additional
perturbative corrections of order $\alpha_s$,
we observe the convergence
between the experimental data and theoretical results obtained
on the basis of approaches combining nonrelativistic QCD and the relativistic quark model.
Our results should be useful for the future comparison with the $(b\bar b)$ and $(c\bar b)$ meson production measurements. Similarily, we
can examine the production rates for doubly heavy mesons including
the $P$- and $D$-wave states. The experimental data for the production
of the mesons $\chi_{c0}+J/\Psi$ are obtained already in 
Refs.\cite{Belle,BaBar}. The work in this direction is in progress.

There is an essential difference in the numerical results
obtained in this work and in Refs.\cite{Ma,BC,BLL} on the basis of
the light-cone formalism. Our results lead to the increase of the
cross section for the production $J/\Psi+\eta_c$ only by a factor
$2\div 2.5$ in the range of center-of-mass energies
$\sqrt{s}=6\div 12$ GeV but not to an order of magnitude at $\sqrt{s}=10.6$ Gev as
in Refs.\cite{Ma,BC,BLL}. While the cross section (24) depends on
the choice of the $c$-quark mass, strong coupling constant
$\alpha_s$, the meson wave functions at the origin,
binding energies $W_{{\cal P,V}}$ and the relativistic
parameter (31), a possible growth of the theoretical value (24) doesn't
solve the problem. So, there exist at least two questions which
could be discussed regarding the performed calculation.

The first question refers to the comparison of our calculation
with the light-cone approach to the same problem in Refs.\cite{Ma,BC,BLL}.
Since the main part of the investigation was connected with the
relative motion of heavy quarks in the meson, it is necessary to
find the wave function $\psi({\bf p_\perp},x)$ which describes bound
states in such an approach and satisfies the quasipotential wave 
equation in the light-front formalism \cite{SW,front,TMF} (see Appendix A).
This equation gives the wave function of a bound state in an arbitrary
Lorentz reference frame. The transformation property for the
wave function from the arbitrary frame to the frame, in which
the total transverse momentum of the two-particle bound state is
equal to zero is determined by Eq.(A3). But contrary to our approach
to the production of doubly heavy mesons, accounting the relative
motion of the heavy quarks, the quark transverse momentum inside
the heavy quarkonium was neglected in Refs.\cite{Ma,BC,BLL}. So,
the basic difference between our method and
the light-cone formalism consists in the fact that in Ref.\cite{Ma,BC,BLL}
the effects of the wave function $\psi(x,{\bf p}_\perp)$ transformation
from the frame ${\bf P}_\perp=0$
to the moving one with the transverse momentum ${\bf P_\perp}\not =0$ were
not taken into account. They lead to terms proportional to the
meson transverse momentum squared ${\bf P}^2_\perp$ in the production
amplitude and might correct the total result.

Finally, another important question regards the accuracy of the calculation performed
in this paper and the numerical estimation of the next term in the expansion
of the production amplitude
over the relative velocity of the fourth order. Recently, corrections
of order $O(v^4)$ were studied in the $S-$ and $P$-wave quarkonium decays in
Refs.\cite{Petrelli,Vairo2006}. There it was demonstrated that
the $v$ expansion converges well for the decays of heavy quarkonium.
Note that a numerical estimation of the parameter $\left\langle{\bf p}^4\right\rangle$
can be obtained as in Eq.(31) by using the dimensional regularization technique
\cite{CMY}: $\left\langle{\bf p}^4/m^4_Q\right\rangle_{DR}$ = 
$\frac{(\tilde W+|B|)^2}{m^2_Q}$. When we consider the production of $1S$-wave charmonium
states with the binding energy $W\ll m_Q$, then the numerical values of the parameters
$\left\langle{\bf p}^2\right\rangle_{DR}$, $\left\langle{\bf p}^4\right\rangle_{DR}$
are small. In the production of excited $S$-wave states the binding energy
increases what leads to the growth of
$\left\langle{\bf p}^2\right\rangle_{DR}$, $\left\langle{\bf p}^4\right\rangle_{DR}$.
So, it follows from the results of Table I that 
$\left\langle{\bf v}^2\right\rangle_{DR}\approx 0.6$ for the $2S$-charmonium states
and $\left\langle{\bf v}^2\right\rangle_{DR}\approx 0.7$ for the $2S$ $(\bar b c)$
meson. Obviously, the convergence of the expansion becomes worse in this case.
The calculation of the coefficient in the term 
$\left\langle{\bf v}^4\right\rangle$ would deserve additional investigation.
On the whole, we estimate the theoretical uncertainty of the obtained production rates due to inaccuracies in the determination
of the meson wave functions at the origin and the omitted terms
in the expansion of the relativistic amplitudes
as the $40\%$ for the $1S$-states and the $70\%$ for the $2S$-states.

\acknowledgments
The authors are grateful to N.Brambilla, K.-T. Chao, R.N.Faustov, V.O.Galkin, A.K.Likhoded,
V.A.Saleev and A.Vairo for general discussions and fruitful remarks.
One of the authors (A.P.M.) thanks M.M\"uller-Preussker and the colleagues from the
Institute of Physics of the Humboldt University in Berlin for warm
hospitality.
The work is performed under the financial support of the
{\it Deutsche Forschungsgemeinschaft} under contract Eb 139/2-3.

\appendix

\section{Quasipotential light-cone distribution amplitude}

In this Appendix we consider the construction of the heavy quark light-cone
distribution amplitude on the basis of the quasipotential equation derived
in light-cone variables.

The quasipotential wave equation for the two-body bound state in the 
light-front formalism
takes the form \cite{SW,front,TMF}:
\begin{equation}
\left[P^2-\frac{\left({\bf p_\perp}+\left(\frac{1}{2}-x\right){\bf P_\perp}
\right)^2+m_1^2}{x}-\frac{\left({\bf p_\perp}+\left(\frac{1}{2}-x\right){\bf P_\perp}
\right)^2+m_2^2}{1-x}\right]\psi_P({\bf p_\perp},x)=
\end{equation}
\begin{displaymath}
=\int_0^1 \frac{dy}{y(1-y)}\int \frac{d{\bf q_\perp}}{2(2\pi)^3}
V(P,x,{\bf p_\perp};y,{\bf q_\perp})\psi_P({\bf q_\perp},y),
\end{displaymath}
where $p$ $({\bf p_\perp})$ and $P$ $({\bf P_\perp})$ are total (transverse) momenta of
relative motion of quarks and the meson. The variable $x$ is introduced
in the following way:
\begin{equation}
x=\frac{1}{2}+\frac{p_+}{P_+},~~p_{\pm}=p_0\pm p_3,~~P_{\pm}=P_0\pm P_3.
\end{equation}
This equation allows to find the two-particle wave function 
in an arbitrary Lorentz reference frame. The transformation of the wave function
from the arbitrary frame to the frame, in which the total transverse momentum
${\bf P}_\perp = 0$ is the following:
\begin{equation}
\psi_P(x,{\bf p_\perp})=\psi_{{\bf P_\perp}=0}(x,{\bf p_\perp}+
(1/2-x){\bf P_\perp}).
\end{equation}
The case of spin particles is discussed in Refs.\cite{TMF,AAH}. In the equal mass case $(m_1=m_2=m)$
in the frame where ${\bf P_\perp}=0$ we get:
\begin{equation}
\left[M^2-\frac{{\bf p_\perp}^2+m^2}{x(1-x)}\right]
\psi({\bf p_\perp},x)=\int_0^1 \frac{dy}{y(1-y)}\int 
\frac{d{\bf q_\perp}}{2(2\pi)^3}V(x,{\bf p_\perp};
y,{\bf q_\perp})\psi({\bf q_\perp},y).
\end{equation}

\begin{figure}[htbp]
\centering
\includegraphics{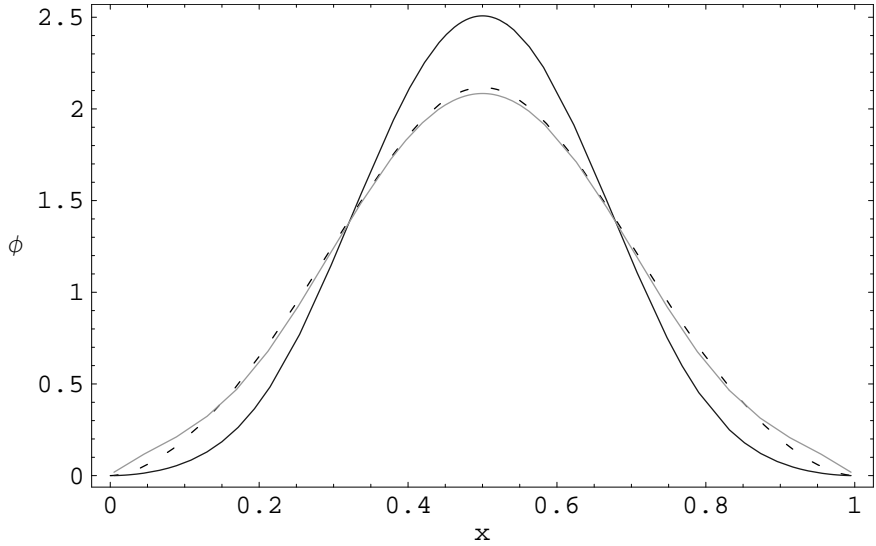}
\vspace{-2mm}
\caption{The light-front distribution amplitude over the momentum fraction $x$
(solid line). The dashed line shows the phenomenological function
from Ref.\cite{BC}. The thin solid line shows the result 
obtained in our model by means of the relations (44), (49) of Ref.\cite{BKL1}
with the c-quark mass $m_c=1.55$ Gev.}
\end{figure}

In order to discuss the realistic wave function $\psi({\bf p_\perp},x)$
there is the need to transform Eq.(A4) to the ordinary form (25), because
we know the potential (29) for it which describes the energy spectrum with sufficiently high accuracy.
Let us introduce the relative momentum ${\bf p}=({\bf p_\perp},p_3)=(p\sin\theta\cos\phi,
p\sin\theta\cos\phi,p\cos\theta)$:
\begin{equation}
p_3=\frac{x-\frac{1}{2}}{\sqrt{x(1-x)}}\sqrt{{\bf p_\perp}^2+m^2},~~
x=\frac{1}{2}\left(1+\frac{p_3}{\epsilon(p)}\right)=\frac{1}{2}\left(1+\frac{p\cos\theta}
{\sqrt{p^2+m^2}}\right),
\end{equation}
so that
\begin{equation}
dx d^2p_\perp=\frac{2x(1-x)}{\sqrt{p^2+m^2}}d{\bf p}.
\end{equation}
As a result, we obtain from Eq.(A4) the ordinary quasipotential equation (25), which contains the wave function $\Psi_0({\bf p})$
connected with the initial one by the following relation:
\begin{equation}
\Psi_0({\bf p})=\left[\frac{4x(1-x)}{{\bf p_\perp}^2+m^2}\right]^{1/4}
\psi({\bf p_\perp},x).
\end{equation}
The ordinary normalization condition for the wave function $\Psi_0({\bf p})$ can
be rewritten for the function $\psi({\bf p}_\perp,x)$ as follows:
\begin{equation}
\int_0^1\frac{dx}{x(1-x)}\int\frac{d{\bf p}_\perp}{2(2\pi)^3}|\psi({\bf p}_\perp,x)|^2=1.
\end{equation}

Then we determine the light-cone distribution amplitude by means of the
quasipotential light-front function $\psi({\bf p}_\perp,x)$:
\begin{equation}
\phi(x)={\cal N}\int \psi({\bf p_\perp},x)d^2{\bf p_\perp}={\cal N}\cdot 8\pi\cdot x(1-x)
\int_{m\sqrt{\frac{(x-1/2)^2}{x(1-x)}}}^\infty
(p^2+m^2)^{1/4} p \Psi_0(p)dp.
\end{equation}
The integral function in
Eq.(A9) differs by the power of the relativistic energy $\epsilon(p)$
($\epsilon^{1/4}(p)\to\epsilon^{1/2}(p)$) and the function $x(1-x)$ from the expression in Ref.\cite{BKL1}.
The normalization factor ${\cal N}$ is determined by the condition:
$\int_0^1\phi(x)dx=1$. The plot of the function $\phi(x)$ and the
comparison with the corresponding amplitudes in Refs.\cite{BC,BKL1}
are presented in Fig.4. All these functions, obtained in the frame
where the meson transverse momentum ${\bf P_\perp}=0$,
have the similar shape. But our function (solid line in Fig.4) has a larger
maximum and falls quicker to the end points $x=0,1$.
To derive the light-cone distribution in the arbitrary reference
frame we can use Eqs.(A3), (A9). The shift of the transverse momentum ${\bf p_\perp}\to {\bf p_\perp}-\left(\frac{1}{2}-x\right){\bf P_\perp}$
shows that the quark light-cone
distribution doesn't change the shape in the arbitrary frame with ${\bf P_\perp}\not =0$.
So, the light-cone amplitude
transformation from dashed line of Ref.\cite{BC} to our
solid line in Fig.4 decreases the production rates of the
charmonium states in the light-front approach.


\begin{thebibliography}{99}
\bibitem{BFY}E.Braaten, S.Fleming, T.C.Yuan, Ann. Rev. Nucl. Part. Sci.
{\bf 46}, 197 (1996).
\bibitem{UFN1}S.S.Gershtein, V.V.Kiselev, A.K.Likhoded, A.V.Tkabladze,
Phys. Usp. {\bf 38}, 1 (1995).
\bibitem{QWG}N.Brambilla, et al. Heavy Quarkonium Physics,
FERMILAB-FN-0779, CERN Yellow Report, CERN-2005-005, e-preprint hep-ph/0412158.
\bibitem{Bali}G.S.Bali, Phys. Rep. {\bf 343}, 1 (2001).
\bibitem{Belle1}K.Abe, et al. Phys. Rev. Lett. {\bf 89}, 142001
(2002).
\bibitem{Belle}K.Abe, et al. Phys. Rev. D {\bf 70}, 071102 (2004).
\bibitem{BaBar}B.Aubert, et al. Phys. Rev. D {\bf 72}, 031101 (2005).
\bibitem{BL1}E.Braaten, J.Lee, Phys. Rev. D {\bf 67}, 054007 (2003); Phys. 
Rev. D {\bf 72}, 099901(E) (2005).
\bibitem{Chao}K.-Y. Liu, Z.-G. He, K.-T. Chao, Phys. Lett. B {\bf 557}, 45 (2003).
\bibitem{Chao1}K.-Y. Liu, Z.-G. He, K.-T. Chao, Search for excited charmonium
states in $e^+e^-$ annihilation at $\sqrt{s}=10.6$ Gev, e-preprint hep-ph/0408141,
v.2.
\bibitem{BBL}G.T.Bodwin, E.Braaten, G.P.Lepage, Phys. Rev. D {\bf 51}, 1125
(1995).
\bibitem{Ma}J.P.Ma, Z.G.Si, Phys. Rev. D {\bf 70}, 074007 (2004).
\bibitem{BC}A.E.Bondar, V.L.Chernyak, Phys. Lett. B {\bf 612}, 215 (2005).
\bibitem{BLL}V.V.Braguta, A.K.Likhoded, A.V.Luchinsky, Phys. Rev.
D {\bf 72}, 074019 (2005).
\bibitem{ZGC}Y.-J. Zhang, Y.-J. Gao, K.-T. Chao, Phys. Rev. Lett.
{\bf 96}, 092001 (2006).
\bibitem{rqm4}D. Ebert, R. N. Faustov, V. O. Galkin, A. P. Martynenko,
Phys. Rev. D  {\bf 70}, 014018 (2004).
\bibitem{apm2005}A.P.Martynenko, Phys. Rev. D {\bf 72}, 074022 (2005).
\bibitem{savrin}V.A.Matveev, A.N.Tavkhelidze, V.I.Savrin, A.N.Sisakian,
Theor. Math. Phys. {\bf 132}, 1119 (2002).
\bibitem{Brodsky}S.J.Brodsky, J.R.Primack, Ann. Phys. {\bf 52}, 315 (1969).
\bibitem{F1973}R.N.Faustov, Ann. Phys. {\bf 78}, 176 (1973).
\bibitem{Form}J.A.M. Vermaseren, FORM, e-preprint math-ph/0010025.
\bibitem{VK}V.V.Kiselev, Int. J. Mod. Phys. A {\bf 10}, 465 (1995).
\bibitem{rqm1}D. Ebert, R. N. Faustov and V. O. Galkin, Phys. Rev. D {\bf 62},
034014 (2000).
\bibitem{rqm2} D. Ebert, R. N. Faustov and V. O. Galkin, Phys. Rev. D
{\bf 67}, 014027 (2003).
\bibitem{rqm3} D. Ebert, R. N. Faustov, V. O. Galkin, A. P. Martynenko,
Phys. Rev. D  {\bf 66}, 014008 (2002).
\bibitem{LSG}W.Lucha, F.F.Sch\"oberl, D.Gromes, Phys. Rep. {\bf 200}, 127
(1991).
\bibitem{CL}W.E.Caswell, G.P.Lepage, Phys. Lett. B {\bf 167}, 437 (1986).
\bibitem{Patrick}P.Labelle, Phys. Rev. D {\bf 58}, 093013 (1998).
\bibitem{RMP}N.Brambilla, A.Pineda, J.Soto, A.Vairo, Rev. Mod.
Phys. {\bf 77}, 1423 (2005).
\bibitem{NR1}B.A.Kniehl, A.A.Penin, V.A.Smirnov, M.Steinhauser, Nucl. Phys. B
{\bf 635}, 357 (2002).
\bibitem{NR2}N.Brambilla, A.Pineda, J.Soto, A.Vairo, Phys. Rev. D
{\bf 60}, 091502 (1999).
\bibitem{Labelle}P.Labelle, G.P.Lepage, U.Magnea, Phys. Rev. Lett. {\bf 72},
2006 (1994).
\bibitem{aKM}W.-Y. Keung, I.J.Muzinich, Phys. Rev. D {\bf 27}, 1518 (1983).
\bibitem{KM}A.I.Milstein, I.B.Khriplovich, JETP {\bf 79}, 379 (1994).
\bibitem{CMY}A.Czarnecki, K.Melnikov, A.Yelkhovsky, Phys. Rev. A {\bf 61},
052502 (2000).
\bibitem{BKL}G.T.Bodwin, D.Kang, J.Lee, Potential-model calculation of an
order-$v^2$ NRQCD matrix element, e-preprint hep-ph/0603186.
\bibitem{MF1999}A.P.Martynenko, R.N.Faustov, JETP, {\bf 88}, 672 (1999).
\bibitem{FFS}P.Falkensteiner, H.Grosse, F.F.Sch\"oberl, P.Hertel,
Comp. Phys. Comm. {\bf 34}, 287 (1985).
\bibitem{W1}B.A.Kniehl, A.A.Penin, Nucl. Phys. B {\bf 577}, 197 (2000).
\bibitem{W2}K.Melnikov, A.Yelkhovsky, Phys. Rev. D {\bf 59}, 114009 (1999).
\bibitem{W3}M.Beneke, A.Signer, V.A.Smirnov, Phys. Lett. B {\bf 454}, 137 (1999).
\bibitem{PDG}Particle Data Group, S.Eidelman et al.,
Phys. Lett.B {\bf 592}, 1 (2004).
\bibitem{SW}S.Weinberg, Phys. Rev. {\bf 150}, 1313 (1966).
\bibitem{front}G.P.Lepage, S.J.Brodsky, Phys. Rev. D {\bf 22}, 2157 (1980).
\bibitem{TMF}V.R.Garsevanishvili, A.N.Kvinikhidze, V.A.Matveev,
A.N.Tavkhelidze, R.N.Faustov,
Theor. Math. Phys. {\bf 23}, 310 (1975).
\bibitem{Petrelli}G.T.Bodwin, A.Petrelli, Phys. Rev. D {\bf 66}, 094011 
(2002).
\bibitem{Vairo2006}N.Brambilla, E.Mereghetti, A.Vairo, Electromagnetic
quarkonium decays at order $v^7$, e-preprint hep-ph/0604190.
\bibitem{AAH}A.A.Khelashvili, JINR Communications P2-8750 (1975).
\bibitem{BKL1}G.T.Bodwin, D.Kang, J.Lee, Reconciling the light-cone
and NRQCD approaches to calculating $e^+e^-\to J/\Psi+\eta_c$,
e-preprint hep-ph/0603185.
\end{thebibliography}
\end{document}